\documentclass[pra,aps,showpacs,superscriptaddress,nofootinbib]{revtex4-1}
\usepackage{graphicx}
\usepackage{dcolumn}
\usepackage{bm}
\usepackage[bbgreekl]{mathbbol}
\usepackage{amsmath}
\usepackage{xr}
\usepackage{color}

\def\bfk{{\bf k}} 
\def\bfq{{\bf q}}
\def\bfp{{\bf p}}
\def\bfr{{\bf r}}
\def\bfR{{\bf R}}
\def\bfE{{\bf E}}

\definecolor{newGreen}{rgb}{0.0, 0.7, 0.0}
\definecolor{newBlue}{rgb}{0.0, 0.4, 1.0}

\newcommand{\arr}[2]
{
\begin{array}{#1}
#2
\end{array}
}
\newcommand{\eq}[1]{
\begin{equation}
\arr{c}{#1}
\end{equation}
}
\newcommand{\vhat}[1]{\hat{\mathbf{#1}}}
\newcommand{\ket}[1]{\left|#1\right\rangle}
\newcommand{\bra}[1]{\langle#1|}
\newcommand{\figref}[1]{Fig. \ref{#1}}
\newcommand{\eqrefTemp}[1]{Eq. \eqref{#1}}
\newcommand{\vecTemp}[1]{\mathbf{#1}}
\newcommand{\Sum}{\displaystyle \sum }
\newcommand{\im}[1]{\text{Im}\;#1}
\newcommand{\re}[1]{\text{Re}\;#1}

\usepackage{comment}

\newcommand{\Int}{\displaystyle \int}

\usepackage[dvipsnames]{xcolor}
\usepackage{hyperref}
\hypersetup{colorlinks=true,
linkcolor = black,
urlcolor  = black,
citecolor = MidnightBlue,
pdfborder={0 0 0}}

\begin{document}
\title{The maximum refractive index of an atomic crystal -- from quantum optics to quantum chemistry}
\author{Francesco Andreoli}
\address{ICFO-Institut de Ciencies Fotoniques, The Barcelona Institute of Science and Technology, 08860 Castelldefels (Barcelona), Spain.}
\author{Bennet Windt}
\address{ICFO-Institut de Ciencies Fotoniques, The Barcelona Institute of Science and Technology, 08860 Castelldefels (Barcelona), Spain.}
\address{Blackett Laboratory, Imperial College London, Prince Consort Road, London SW7 2AZ, UK}
\address{Max Planck Institute of Quantum Optics, Hans-Kopfermann-Stra{\ss}e 1, 85748 Garching, Germany}
\author{Stefano Grava}
\address{ICFO-Institut de Ciencies Fotoniques, The Barcelona Institute of Science and Technology, 08860 Castelldefels (Barcelona), Spain.}
\author{Gian Marcello Andolina}
\address{ICFO-Institut de Ciencies Fotoniques, The Barcelona Institute of Science and Technology, 08860 Castelldefels (Barcelona), Spain.}
\author{Michael J. Gullans}
\address{Joint Quantum Institute, NIST/University of Maryland, College Park, Maryland 20742, USA}
\address{Joint Center for Quantum Information and Computer Science, NIST/University of Maryland, College Park, Maryland 20742, USA}
\author{Alexander A. High}
\address{Pritzker School of Molecular Engineering, University of Chicago, Chicago, Illinois 60637, USA}
\address{Center for Molecular Engineering and Materials Science Division, Argonne National Laboratory, Lemont, Illinois 60439, USA}
\author{Darrick E. Chang}
\address{ICFO-Institut de Ciencies Fotoniques, The Barcelona Institute of Science and Technology, 08860 Castelldefels (Barcelona), Spain.}
\address{ICREA-Instituci{\'o} Catalana de Recerca i Estudis Avan{\c c}ats, 08015 Barcelona, Spain}

\date{\today}

\begin{abstract}
All known optical materials have an index of refraction of order unity. Despite the tremendous implications that an ultrahigh index material could have for optical technologies, little research has been done on why the refractive index of materials is universally small, and whether this observation is truly fundamental. Here, we describe significant insights that can be made, by posing and quantitatively analyzing a slightly different problem -- what is the largest refractive index that one might expect from an ordered arrangement~(crystal) of atoms, as a function of atomic density. At dilute densities, this problem falls into the realm of quantum optics, where atoms do not interact with one another except via the scattering of light. On the other hand, when the lattice constant $d\sim a_0$ becomes comparable to the Bohr radius, the electronic orbitals centered on different nuclei begin to overlap and strongly interact, giving rise to quantum chemistry. We present a minimal model that allows for a unifying theory of index spanning the quantum optics and quantum chemistry regimes. A key aspect of this theory is its treatment of multiple light scattering, which can be highly non-perturbative over a large density range, and which is the reason that conventional theories to predict the refractive index break down. In the quantum optics regime, we show that ideal light-matter interactions can have a single-mode nature, which allows for a purely real refractive index that grows with density as $(N/V)^{1/3}$. At the onset of quantum chemistry, we show how two physical mechanisms -- excited electron tunneling dynamics and the buildup of ground-state density-density correlations -- play dominant roles in opening up inelastic or spatial multi-mode light scattering processes, which ultimately reduce the index back to order unity while simultaneously introducing absorption. Around the onset of chemistry, our theory predicts that ultrahigh index~($n\sim 30$), low-loss materials could in principle be allowed by the laws of nature. This work could inspire new efforts to design and realize materials with ultrahigh index, and also stimulate the investigation of other exotic physics driven by the interplay of collective optical phenomena, multiple scattering, and quantum chemistry.
\end{abstract}

\maketitle


\section{Introduction}
Ultrahigh index, low-loss optical materials at telecom or visible frequencies would have potentially game-changing implications for optical technologies and light-based applications. Notably, the reduction in the optical wavelength $\lambda=\lambda_0/n$ compared to the free-space value $\lambda_0$ could lead to unprecedented opportunities associated with field confinement and focusing. For example, nanoscale resonators and waveguides could lead to strong optical nonlinear interactions at the single-photon level, compact metasurfaces~\cite{yu_flat_2013,zou_dielectric_2013,yu_flat_2014,khorasaninejad_metalenses_2016} such as for wavefront shaping, and optical circuitry with the same physical footprint as electronic transistors. The reduction in wavelength could also be useful for nanoscale microscopy and optical lithography~\cite{lin_optical_2006}. Despite these potential implications, known optical materials at telecom and visible wavelengths ubiquitously have an index of order unity, and limited research has been done on why such a limitation might arise~\cite{khurgin_expanding_2022,shim_fundamental_2021}. For example, using Kramers-Kronig relations and a sum rule, one recent work places an elegant bound between the maximum index a transparent material can have and the bandwidth over which it can be sustained~\cite{shim_fundamental_2021}, but does not directly address how large the index can be. Here, we introduce a minimal physical model that elucidates how large we \textit{might expect} the index to become, under ideal circumstances, and the fundamental mechanisms that limit its indefinite growth. Our analysis is limited to optical frequencies, with the specific assumption that only the electronic response contributes to the refractive index.

\begin{figure}[t]
\centering
\includegraphics[width=0.6\textwidth]{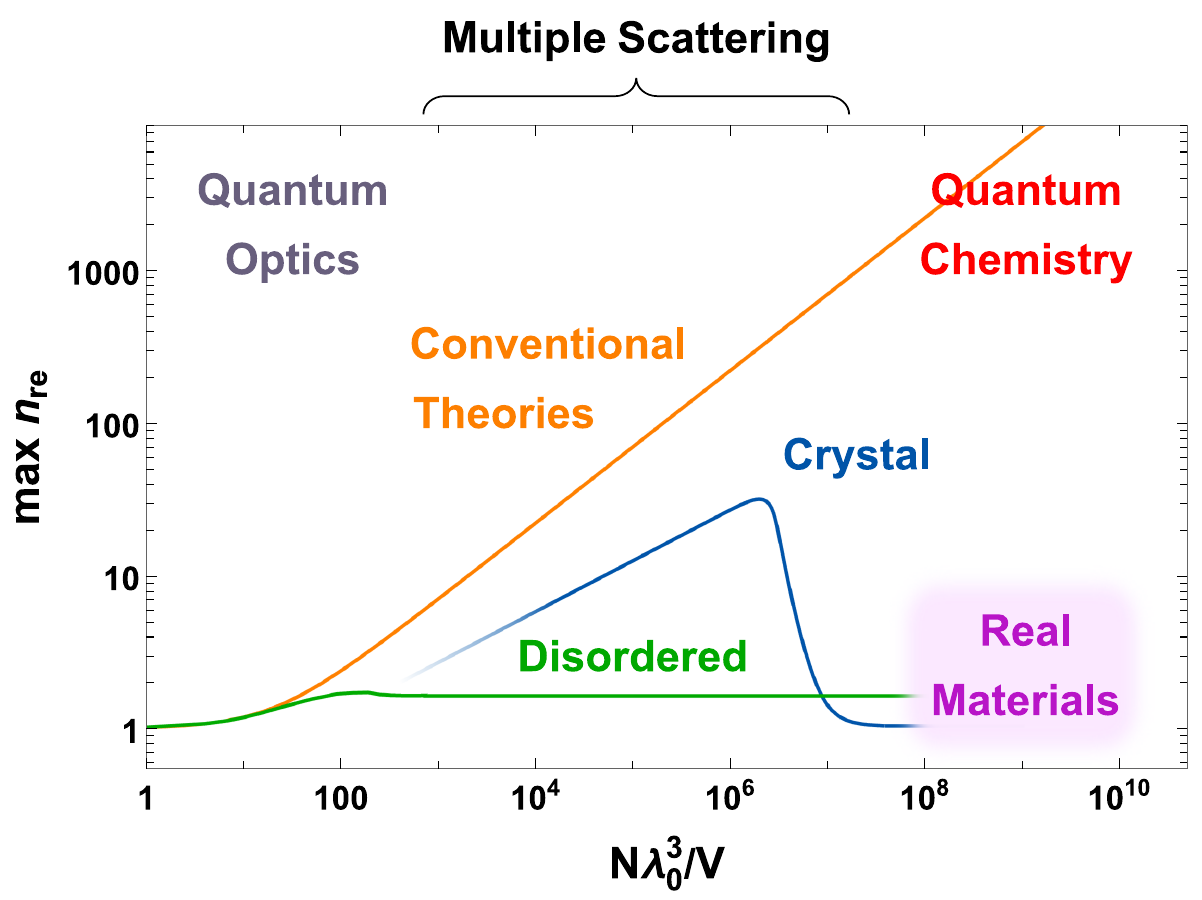}
\caption{
Schematic plot of the maximum attainable real part of the refractive index versus density of atoms $N/V$~(in units of the wavelength $\lambda_0$ of an atomic transition). Conventional theories predict a maximum index that scales with density as $n\sim (N\lambda_0^3/V)^{1/2}$~(orange curve) near an atomic resonance. In the ``quantum optics'' regime, atoms are sufficiently far enough part that they can be considered isolated objects, which only interact via the electromagnetic field. For sufficiently high densities, one enters the ``quantum chemistry'' regime where the overlap of electronic orbitals between different atoms becomes non-negligible and chemical interactions occur. In a vast intermediate regime of densities, ranging from $N\lambda_0^3/V\sim 1$ through the onset of quantum chemistry, a non-perturbative treatment of multiple scattering of light is needed to correctly predict the index. For a disordered atomic medium in the quantum optics regime, it was found that the maximum index is limited to $n\approx 1.7$~(green curve)~\cite{andreoli_maximum_2021}. In this work, we show that a perfect crystal exhibits a purely real index scaling as $n\sim (N\lambda_0^3/V)^{1/3}$ in the quantum optics regime, and the maximum attainable value is limited by effects arising from quantum chemistry~(blue curve).
}
\label{fig:schematic}
\end{figure}

As a starting point for a bottom-up model, we observe that the basic building block of a material -- individual atoms -- can have an extraordinarily large and universal response to light when isolated. In particular, it is well-known that an isolated atom can exhibit a scattering cross section of $\sim\lambda_0^2$ when illuminated by photons resonant with an electronic transition of wavelength $\lambda_0$. Given that a typical transition wavelength $\lambda_0\sim 1\;\mu$m is much larger than the typical spacing between atoms in a solid~(as characterized by the Bohr radius, $a_0\sim 0.05$~nm), one might wonder why the large atomic density in a solid does not provide a strongly multiplicative response to light. Indeed, such a response is predicted by conventional macroscopic theories of the refractive index, such as the Drude-Lorentz, Maxwell-Bloch, or Lorentz-Lorenz models. Specifically, these theories state that the macroscopic index should depend on the product of the polarizability of a single atom and the particle density as $n(\omega)\sim \sqrt{\alpha(\omega)N/V}$~(the Lorentz-Lorenz equation is different, but the conclusion that the maximum index can scale as $(N/V)^{1/2}$ is the same~\cite{andreoli_maximum_2021}). The large near-resonant polarizability of an individual atom then leads to a predicted maximum index of $n\sim 10^5$ at solid densities, as illustrated by the orange curve in Fig.~\ref{fig:schematic}. This is hard to reconcile with empirical observations.

The main goal of this work is to elucidate what the unifying curve in Fig.~\ref{fig:schematic} should look like, i.e. what maximum index might be achievable as a function of atomic density, particularly when the atoms form a perfect crystal~(blue curve). Our theoretical analysis aims to connect two quite different regimes. In particular, at low densities~(which we term the ``quantum optics'' regime), the atomic nuclei are too far separated for electrons centered on different nuclei to directly interact. Then, the atoms only interact via electromagnetic fields, and the index should solely be a function of the lattice constant and the single-atom polarizability. In the ``quantum chemistry'' regime, the atomic densities are sufficiently high that electronic orbitals on neighboring nuclei begin to overlap, in principle giving rise to a wealth of new phenomena associated with chemical interactions or solid-state physics, and ultimately resulting in the formation of a solid. At very dilute densities in the quantum optics regime, one expects that conventional macroscopic theories of index hold. Likewise, empirically we know that computational quantum chemistry can predict the optical properties of real solids with reasonable accuracy. In these limits, the multiple scattering of light is a weak effect. The challenge of constructing a unifying curve in Fig.~\ref{fig:schematic} lies in a vast intermediate scale of densities, starting from a minimum density of $N/V\sim 1/\lambda_0^3$ and spanning through part of the quantum chemistry regime. In this range, the spatial extent of the scattering cross section of an~(isolated) atom can far exceed the inter-atomic distance. Then, multiple scattering of light can become very strong, and in fact causes the breakdown of conventional theories of refractive index. Our model treats multiple scattering non-perturbatively, including in the presence of the onset of quantum chemistry. For context, we note that a complementary part of this puzzle was addressed in Ref.~\cite{andreoli_maximum_2021}~(also see previous historical work~\cite{chomaz_absorption_2012,jenkins_collective_2016,jennewein_coherent_2016,corman_transmission_2017}). In particular, for a completely disordered medium in the quantum optics regime, it was found that the maximum index saturates at $n\approx 1.7$ regardless of how high the atomic density becomes~(solid green curve of Fig.~\ref{fig:schematic}). In Ref.~\cite{andreoli_maximum_2021}, strong disorder renormalization group theory was used to treat multiple scattering non-perturbatively and identify the physical mechanism~(strong, random-strength near-field interactions between a given atom and its single nearest neighbor) by which the refractive index saturates to a maximum value.

We now summarize the scope and main results of the paper. In Sec.~\ref{sec:QO}, we analyze the refractive index of an atomic crystal in the quantum optics limit. We first review a result that has gained theoretical~\cite{bettles_enhanced_2016,shahmoon_cooperative_2017} and experimental~\cite{rui_subradiant_2020} interest in recent years, that a single two-dimensional~(2D) array of atoms can provide a large, lossless and cooperatively enhanced response to light near resonance, as characterized by large reflectance and large phase shift in transmission. By considering a three-dimensional~(3D) crystal as a sequence of 2D arrays separated by lattice constant $d_z$, we then show that the 2D properties directly translate into a refractive index near resonance that can be purely real, and which scales as $n_{\rm max}\propto \lambda_0/d_z$. The key property enabling this behavior is the \textit{single-mode nature} of the light-matter interaction, both in the 2D and 3D arrays, where light excites only a single collective mode of the atoms, \textit{and} this collective mode only re-radiates light elastically back in the same direction, to produce a maximal and lossless response.

We then briefly introduce the model to incorporate quantum chemistry effects. It is well-known that the many-electron problem of computational quantum chemistry for real solids is a challenging and likely intractable problem to solve exactly. State-of-the-art computational techniques, like modern density functional theory, are largely based upon sophisticated approximation techniques. Here, we favor an approach that is less dependent on such approximations. In particular, we limit our theory to the onset of quantum chemistry, or an expansion around a large lattice constant compared to the Bohr radius, $d/a_0$. Then, quantum chemistry can be treated perturbatively, while multiple scattering can still be treated non-perturbatively. This large $d/a_0$ expansion of quantum chemistry and the resulting minimal model is summarized in Sec.~\ref{sec:model}, while a more detailed discussion and derivation is provided in Sec.~\ref{sec:derivation}.

In Sec.~\ref{sec:indexchemistry}, we incorporate the results of quantum chemistry into multiple scattering. Perhaps not surprisingly, effects associated with chemistry can break the single-mode nature of atom-light interactions found in the quantum optics regime, either by allowing for spatial multi-mode response or inelastic light scattering. Considering the simplest model of a lattice of hydrogen atoms, we show that a combination of the emergence of quantum magnetism, electronic density-density correlations, and tunneling dynamics of photo-excited electrons are the primary mechanisms for multi-mode and inelastic scattering at large $d/a_0$. We quantify how these effects lead to a maximum allowed real part of the refractive index, and the growth of the imaginary part associated with absorption. Our model suggests that an ultrahigh index material of $n\sim 30$ with low losses is not fundamentally forbidden by the laws of nature. Although our quantitative model deals with hydrogen atoms, we also discuss possible realistic routes toward ultrahigh-index materials, such as high-density arrays of solid-state quantum emitters or van der Waals heterostructures, and qualitatively show that the ultrahigh index is robust to some degree of additional imperfections (e.g., implementation-dependent inhomogeneities, or additional inelastic mechanisms). In Sec.~\ref{sec:outlook}, we provide an outlook of future interesting research questions to explore.

\section{Refractive index: the quantum optics limit}\label{sec:QO}

\label{sec:index}

\subsection{Formalism}
\label{subsec:Formalism}

In this section, we derive the refractive index of a perfect atomic lattice in the quantum optics limit, where quantum chemistry interactions between atoms are ignored and each atom is seen as a point dipole from the standpoint of its optical properties. Specifically, we consider the relevant levels of the atom to consist of an electronic ground state and first excited state, which are connected by an electric dipole transition of frequency $\omega_0$ and corresponding wavevector $k_0=\omega_0/c$ and wavelength $\lambda_0=2\pi/k_0$. The atoms can also be driven by a weak coherent input field of frequency $\omega_L$, with a polarization $\vhat x$ that aligns with the dipole matrix element $\vecTemp p_0 = p_0 \vhat x$ of the atomic transition. The excited state can only decay by emitting a photon and returning to the ground state, which occurs at a rate $\Gamma_0=k_0^3 p_0^2/3\pi\epsilon_0\hbar $ for an isolated atom. 

Although our conclusions in this section will be completely general to any atom with the properties specified above, here we adopt a second-quantized notation consistent with our later model including quantum chemistry, when we consider a hydrogen atom whose ground and excited states are then the 1s and 2p${}_x$ orbitals. In a rotating frame relative to the driving field and in the long-wavelength limit, the Hamiltonian describing the atom-light interactions is given by~\cite{gross_superradiance:_1982,dung_resonant_2002,asenjo-garcia_exponential_2017}
\begin{eqnarray}
\label{eq:3DQO_Hamiltonian}
    H_{\rm QO} & = & H_0+H_{\rm dip-dip}+H_{\rm drive} \nonumber \\ & = & -\sum_{i\sigma}\delta b^{\dagger}_{pi\sigma}b_{pi\sigma}-\Gamma_0
    \sum_{ij\sigma\sigma'}G_{ij}(b_{pi\sigma}^{\dagger}b_{si\sigma})(b^{\dagger}_{sj\sigma'}b_{pj\sigma'})-\sum_{i\sigma}\left[\Omega_i b^{\dagger}_{pi\sigma}b_{si\sigma}+h.c.\right].
\end{eqnarray}
Here, we have defined the detuning $\delta=\omega_L-\omega_0$, the Rabi frequency $\Omega_i =\vecTemp p_0 \cdot \bm{\mathcal{E}}_{\text{in}}({\bf R}_i)/\hbar$ associated with the coherent input field $\bm{\mathcal{E}}_{\text{in}}(\bfr)$, and the fermionic operator $b_{\alpha i \sigma}$ that annihilates an electron of orbital $\alpha$ and spin $\sigma$ on atom $i$, whose nucleus is at position ${\bf R}_i$. The dipole-dipole interaction $H_{\rm dip-dip}$ describes the electronic excitation $(b_{pi\sigma}^{\dagger}b_{si\sigma})$ of an atom from its s to its p-orbital at site $i$, and the de-excitation $(b^{\dagger}_{sj\sigma'}b_{pj\sigma'})$ of another at site $j$. This captures electromagnetic field mediated interactions once the photons are integrated out within the Born-Markov approximation, with  $G_{ij}=\vhat{ x }\cdot {\bf G}({\bf R}_i-{\bf R}_j,\omega_0)\cdot\vhat{x}$ being proportional to the electromagnetic Green's function at frequency $\omega_0$~(see below). The positive-frequency component of the electric field operator within the same limit is~\cite{asenjo-garcia_exponential_2017}
\begin{eqnarray}
    \vecTemp{E}(\bfr)=\vecTemp{E}_{\rm in}(\bfr)+
    \dfrac{k_0^3}{3\pi\epsilon_0}
    \sum_{i\sigma}\vecTemp{G}(\bfr-{\bf R}_i,\omega_0)\cdot\vecTemp{p}_{0}\;b_{si\sigma}^{\dagger}b_{pi\sigma}, \label{eq:input-output_full}
\end{eqnarray}
which formally expresses the total field at any spatial point, in terms of the input and the field scattered by the atoms. Here, we use the notation $\bm{\mathcal{E}}(\bfr)$ to denote a classical field~(i.e. coherent state) value, while $\vecTemp{E}(\bfr)$ denotes a quantum field operator. At this level of discussion, $H_{\rm QO}$ and the relevant Hilbert space can just as well be written in terms of pseudospin-$1/2$ operators to describe the two-level atoms, as common in quantum optics literature~\cite{gross_superradiance:_1982,dung_resonant_2002,asenjo-garcia_exponential_2017}. We avoid that here, to prevent confusion with the actual electronic spins $\sigma$ and to more naturally extend to the inclusion of quantum chemistry.

The function ${\bf G}(\bfr-\bfr',\omega_0)$ is a dimensionless tensor describing the field at position $\bfr$ radiated by a point dipole at position $\bfr'$ oscillating at frequency $\omega_0$. Its projection onto the $\hat{\bm{x}}$ direction~(giving the $\hat{\bm{x}}$ component of the field radiated by a dipole oriented along $\hat{\bm{x}}$) is explicitly given by 
\begin{equation}
    \vhat x\cdot {\bf G}(\bfr,\omega_0)\cdot \vhat{x}=
\dfrac{3}{4}
e^{i k_0 |\bfr|}\left[\left(\dfrac{1}{k_0 |\bfr|}+\dfrac{i}{(k_0 |\bfr|)^2}-\dfrac{1}{(k_0 |\bfr|)^3}\right)
+\left(-\dfrac{1}{k_0 |\bfr|}-\dfrac{3i}{(k_0 |\bfr|)^2}+\dfrac{3}{(k_0 |\bfr|)^3}\right)\dfrac{ (\vhat x\cdot \bfr)^2  }{|\bfr|^2}\right],
\label{eq:Green tensor}
\end{equation}
Since $G_{ij}$ is complex, the Hamiltonian $H_{\rm dip-dip}$ is non-Hermitian. Its Hermitian and non-Hermitian components describe coherent energy exchange between atoms, and collective spontaneous emission arising from interference of light emission, respectively. To the extent that Eqs.~(\ref{eq:3DQO_Hamiltonian}) and~(\ref{eq:input-output_full}) can be solved exactly, they fully incorporate the effects of non-perturbative multiple scattering of light and wave interference in emission. Derivations of these equations can be found, for example, in Refs.~\cite{gross_superradiance:_1982,dung_resonant_2002,asenjo-garcia_exponential_2017}, and also in Sec.~\ref{sec:derivation} in a manner consistent with our quantum chemistry discussion.

\begin{figure}[t!]
\centering
\includegraphics[width=\textwidth]{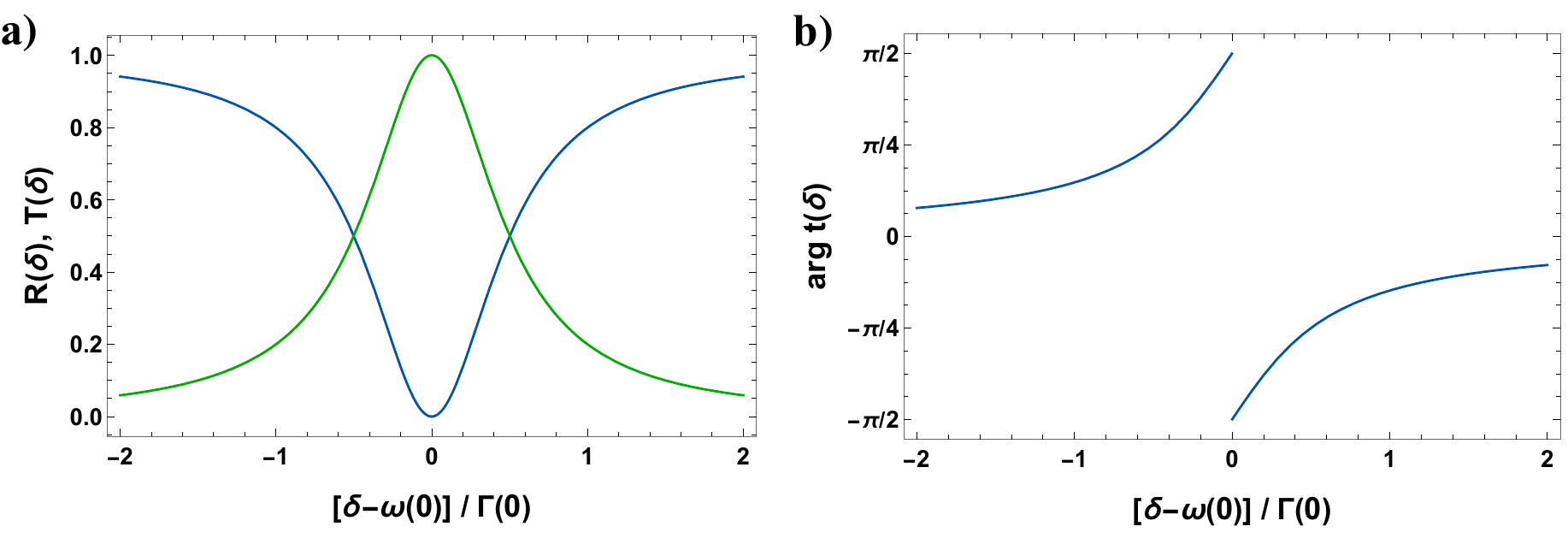}
\caption{\textbf{Reflection and transmission coefficients of a 2D atomic array.} a) Spectrum of reflectance~(blue) and transmittance~(green) as a function of detuning relative to the collective resonance frequency of the array~$\delta-\omega(0)$, and in units of the collective linewidth $\Gamma(0)$. b) Spectrum of transmission phase, ${\rm arg}\; t(\delta)$. 
}
\label{fig:plot_t_r_2D}
\end{figure}

\subsection{Optical response of a 2D array}
\label{subsec:optical_2Darray}

While one can in principle directly study the optical response of a 3D lattice~\cite{de_vries_point_1998,antezza_fano-hopfield_2009}, one can arrive at a better physical understanding of the refractive index by first considering a single, 2D square array of lattice constant $d$, located in the $z=0$ plane. This brief review of a 2D array closely follows the discussions of Refs.~\cite{bettles_enhanced_2016,shahmoon_cooperative_2017}.

We write the total wave function $|\Psi(t)\rangle=|\psi_{\rm 2D}(t)\rangle\otimes |\sigma\rangle$ in terms of the orbital $|\psi_{\rm 2D}(t)\rangle$ and electronic spin $|\sigma\rangle$ wave functions, the latter of which is time independent and irrelevant in the quantum optics limit, as atom-light interactions and thus $H_{\rm QO}$ are decoupled from spin. In the single-excitation limit~(containing exactly one p-orbital), the discrete translational symmetry implies that all eigenstates of $H_{\rm dip-dip}$ of the 2D array are Bloch modes with corresponding Bloch wavevector ${\bf k}_{xy}$, $|E_{{\bf k}_{xy}}\rangle=N^{-1/2}\sum_i e^{i\bfk_{xy}\cdot \bfR_i} b_{pi}^{\dagger}b_{si}|G\rangle$. The ground state consists of all atoms in the s orbitals, $|G\rangle=\Pi_i b_{si}^{\dagger}|{\rm vac}\rangle$. Here, we have suppressed the spin index given its decoupling from dynamics, and $N\rightarrow \infty$ represents the number of atoms in the 2D array. We write the complex eigenvalues of the Bloch modes in the form $\omega(\bfk_{xy})-i\Gamma(\bfk_{xy})/2=-\Gamma_0
\sum_{i,j} G_{ij} e^{i \bfk_{xy} \cdot {\bfR}_{ij}}$, which can be calculated by discrete Fourier transform of the Green's function~\cite{bettles_enhanced_2016,shahmoon_cooperative_2017}. The dispersion relation $\omega(\bfk_{xy})$ represents the energy shift of each Bloch mode relative to the bare atomic resonance $\omega_0$, due to dipole-dipole interactions, and can be evaluated numerically. The collective emission rate admits the analytic solution $\Gamma(\bfk_{xy})= [3\lambda_0^2\Gamma_0/(4\pi d^2)] \Theta(k_0-|\bfk_{xy}|) (k_0^2-k_x^2)/(k_0\sqrt{k_0^2-|\bfk_{xy}|^2})$ (where $\Theta(k_0-|\bfk_{xy}|)$ is the Heaviside step function) when the lattice constant $d<\lambda_0/2$, and is modified from the single-atom value due to interference in the emitted light from different atoms~\cite{shahmoon_cooperative_2017}. For a collective mode with uniform phase~($\bfk_{xy}=0$), one has $\Gamma(0)=3\lambda_0^2\Gamma_0/4\pi d^2$. In particular, for small lattice constants, the rate is significantly enhanced relative to $\Gamma_0$ by an amount $\propto (\lambda_0/d)^2$ due to strong constructive interference. On the other hand, when $|\bfk_{xy}|>k_0$, the modes become perfectly subradiant, $\Gamma(\bfk_{xy})=0$, due to an impedance mismatch between the wavevector of the excitation and the dispersion relation of propagating light.

We now consider driving with a plane wave at normal incidence to the 2D array~(with longitudinal wavevector $k_z=k_0$ and perpendicular wavevector $\bfk_{xy}=0$), whose spatially uniform Rabi frequency $\Omega_i=\Omega_0$ is sufficiently weak that dynamics can be restricted to the ground state and single-excitation manifold. The discrete symmetry imposes that this field will only couple to the Bloch mode $|E_{\bfk_{xy}=0}\rangle$, with the time-dependent wave function restricted to the form $|\psi_{\rm 2D}(t)\rangle=c_G(t)|G\rangle+c_E(t)|E_{\bfk_{xy}=0}\rangle$. The wave function approach to the non-Hermitian Hamiltonian~(\ref{eq:3DQO_Hamiltonian}), or more properly the full master equation, is valid within the quantum jump formalism of open systems. Furthermore, under weak driving, quantum jumps can be neglected and $c_G(t)\approx 1$ up to order $(\Omega_0/\Gamma_0)^2$~\cite{manzoni_optimization_2018}. The Schrodinger equation then leads to a steady-state amplitude of the excited state whose dependence on detuning $\delta$ goes as
\begin{equation}
    c_E(\delta)=\dfrac{\Omega_0}{-\delta+\omega(0)-i\Gamma(0)/2}. 
    \label{eq:cE}
\end{equation}

We now derive the expectation value $\bm{\mathcal{E}} (\bfr)=\langle \bfE(\bfr)\rangle$ of the total field from Eq.~(\ref{eq:input-output_full}). Given the periodic nature of the array and that only the  $\bfk_{xy}=0$ Bloch mode is excited, the total field only contains transverse momentum components given by integer multiples $(m,n)$ of the reciprocal lattice vectors, ${\bf g}_{mn}=(2\pi/d)(m\hat{\bm{x}}+n\hat{y})$. Specifically, we find 
\eq{
    \bm{\mathcal{E}}(\bfr)=  \bm{\mathcal{E}}_{\text{in}}(\bfr) +
    \mathcal{E}_{0}
    \left(
    i\dfrac{ \Gamma(0)}{2}
    \displaystyle \sum_{m,n} 
    \dfrac{
    k_0^2-( {\bf g}_{mn}\cdot \vhat x)^2
    }{k_0k_{z}^{(m,n)}}
    e^{i{\bf g}_{mn}\cdot \bfr_{\perp}+ik_{z}^{(m,n)} |z|} 
    \right) \dfrac{c_E(\delta)}{\Omega_0}
    ,
    \label{eq:diffracted_field}
}
where $k_{z}^{(m,n)}=\sqrt{k_0^2-|{\bf g}_{mn}|^2}$, and where $\bm{\mathcal{E}}_{\text{in}}(\bfr) =\mathcal E_0 e^{ik_0 z}\vhat x$. Note that for $d<\lambda_0$, $k_z^{(m,n)}$ is imaginary except for $m=n=0$. In other words, only transmission and reflection at normal incidence are radiation waves, while any other $(m,n)\neq (0,0)$ correspond to evanescent diffraction orders~(with the sign of $\textrm{Im}\;k_z$ chosen such that the field decays away from the array). In the far field limit~(large $|z|$), one thus has
\eq{
\label{eq:diffracted_field_subwavelength}
   \bm{\mathcal{E}}(\bfr_{\perp}, |z|\gg 1/k_0) \approx 
    \bm{\mathcal{E}}_{\text{in}}(\bfr) \left[1+i\dfrac{\Gamma(0)}{2}\dfrac{c_E(\delta)}{\Omega_0}\Theta(z)\right] +
     \bm{\mathcal{E}}_{\text{in}}^*(\bfr) \left[i\dfrac{\Gamma(0)}{2}\dfrac{c_E(\delta)}{\Omega_0}\Theta(-z)\right],
}
Using the steady-state amplitude in Eq.~(\ref{eq:cE}), we identify the reflection and transmission coefficients $r(\delta)= i\Gamma(0) /[-2\delta+2\omega(0)-i\Gamma(0)]$ and $t(\delta)=1+r(\delta)$. Note in particular that the array is perfectly reflecting when light is resonant with the Bloch mode, $\delta=\omega(0)$, and generally that the system is lossless with $|r|^2+|t|^2=1$~\cite{bettles_enhanced_2016,shahmoon_cooperative_2017}. These properties reflect the \textit{single-mode nature} of the light-matter interaction for this system, where the light excites only a single collective eigenmode $|E_{\bfk_{xy}=0}\rangle$, \textit{and} this collective mode only re-radiates light elastically back in the same $\bfk_{xy}=0$ direction~(either forward or backward). In Fig. \ref{fig:plot_t_r_2D}, we plot the reflectance and transmittance spectra and the transmission phase. Notably, near resonance, the transmitted light can undergo a significant phase shift of up to $\pi/2$.

\begin{figure}[t!]
\centering
\includegraphics[width=0.5\textwidth]{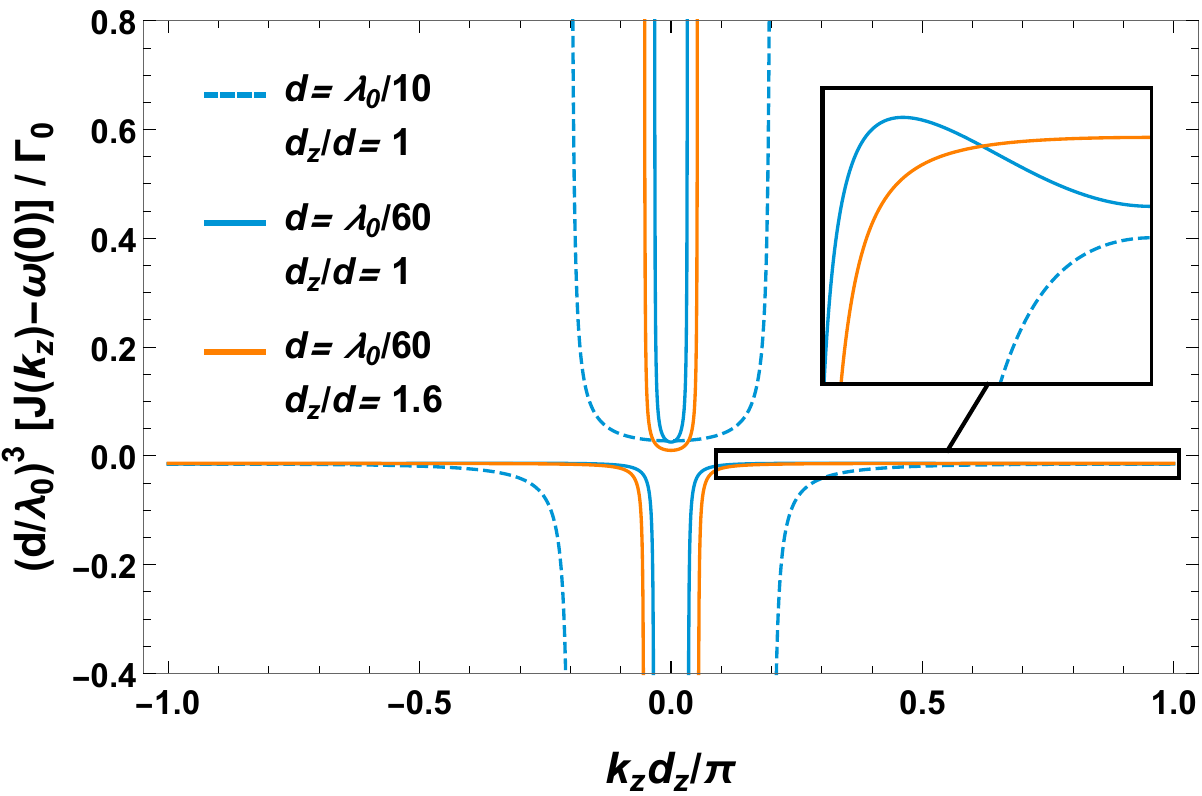}
\caption{\textbf{Optical band structure along $\vhat z$ of a 3D lattice.} Re-scaled dispersion relation $(d/\lambda_0)^3 J(k_z)/\Gamma_0$ for Bloch waves along $z$ with Bloch wavevector $k_z$, for several different lattice constants $d$ and $d_z$ indicated by the legend. The non-invertibility of the band for some values of $d,d_z$ can be clearly seen in the inset. 
}
\label{fig:fra_band_and_scaling}
\end{figure}


\subsection{Refractive index}\label{subsec:indexQO}

We now consider a 3D array, with the lattice constant $d_z$ between 2D layers allowed to be different than the intra-layer lattice constant $d$. Naively, if each 2D layer can contribute a large phase shift to transmitted light, then one expects a large, perfectly real index scaling like $n\sim \lambda_0/d_z$. This naive argument does not account for multiple scattering between planes or evanescent fields, but we now present an exact calculation showing that this scaling holds.

As before, we restrict ourselves to the weak driving limit at normal incidence. Thus, only the collective mode $|E_{\bfk_{xy}=0}\rangle$ of each 2D array can be excited, leading to a total wave function $|\psi_{\rm 3D}(t)\rangle=c_G(t)|G\rangle+\sum_j c_E^{j}(t)|E_{\bfk_{xy}=0,j}\rangle$, where $c_G(t)\approx 1$ and $|E_{\bfk_{xy}=0,j}\rangle$ is the collective mode associated with the 2D plane at position $z_j$. Within this manifold, the dynamics under $H_{\rm QO}$ of Eq.~(\ref{eq:3DQO_Hamiltonian}) is equivalent to a 1D problem, characterized by the matrix elements 
$H^{\rm 1D}_{{\rm dip-dip},ij}=\langle E_{\bfk_{xy}=0,i}|H_{\rm dip-dip}|E_{\bfk_{xy}=0,j}\rangle$, which explicitly read
\begin{equation} 
H^{\rm 1D}_{{\rm dip-dip},ij}=
\left\{
\begin{array}{ll}
\omega (0) - \dfrac{i}{2}  \Gamma \left(0\right) &\;\;\;\;\;\;\;\;i= j,
\\\\
    -i\dfrac{ \Gamma(0)}{2}
    \displaystyle \sum_{m,n} 
    \dfrac{
    k_0^2-( {\bf g}_{mn}\cdot \vhat x)^2
    }{k_0k_{z}^{(m,n)}}
    e^{ik_{z}^{(m,n)} |z_i-z_j|} &\;\;\;\;\;\;\;\;i\neq j.
    \end{array}
    \right.\label{eq:H1Ddip-dip}
\end{equation}
Comparing with Eq.~(\ref{eq:diffracted_field}), the off-diagonal elements $i\neq j$ between different planes can equivalently be interpreted as the Rabi frequency associated with the field scattered by one plane, as experienced by atoms in another plane. 

Diagonalizing this matrix $H^{\rm 1D}_{{\rm dip-dip}}$ then gives the optical band structure of the array at normal incidence, with dispersion relation 
\eq{
\label{eq:optical_band_equation}
J(k_z) =\omega(0)+
 \dfrac{\Gamma(0)}{2}\left[ \dfrac{\sin(k_0 d_z)}{\cos(k_z d_z)-\cos( k_0 d_z)}  + J_{\text{ev}}(k_z)\right],
}
where $|k_z|\leq \pi/d_z$ is restricted to the first Brillouin zone. Recall that we are in a rotating frame, and $J(k_z)$ is thus the frequency relative to the bare atomic resonance frequency $\omega_0$. Also, we will always be in a regime where the shift is small relative to the bare frequency $|J(k_z)|/\omega_0 \ll 1$. Here, $J_{\rm ev}(k_z)$ is the contribution coming from the evanescent fields of each plane, and is found to be
\eq{
\label{eq:Jev_full_equation}
J_{\text{ev}}(k_z)=-
\displaystyle \sum_{(m,n)\neq(0,0)} 
\dfrac{
({\bf g}_{mn}\cdot \vhat x/k_0)^2-1
}{\sqrt{\left|{\bf g}_{mn}/k_0\right|^2-1}}
 \left[ 1 + \dfrac{\sinh\left( k_0  d_{z} \sqrt{\left|{\bf g}_{mn} /k_0\right|^2-1}\right)}{\cos\left(k_z d_z\right) - \cosh\left( k_0  d_{z} \sqrt{\left|{\bf g}_{mn} /k_0\right|^2-1}\right)}\right].}
Although $H^{\rm 1D}_{{\rm dip-dip}}$ itself is non-Hermitian, the dispersion relation is purely real, as a result of the lossless nature of the individual planes.

A typical band structure is illustrated in Fig. \ref{fig:fra_band_and_scaling} for several different values of $d/\lambda_0$ and $d_z/d$. As long as $J(k_z)$ is invertible~(a single value of $|k_z|$ is associated to each value of $J$), then the index is well-defined and $J(k_z)$ can be used to directly infer its value. In particular, we consider the band edge $k_z=\pi/d_z$ and use the fact that $|J(k_z)|/\omega_0 \ll 1$. Then, the maximum index, describing the reduction of the effective wavelength of light compared to free space at the same frequency, is (in the relevant regime of $d<\lambda_0/2$)
\eq{
\label{eq:fra_max_index_invertible}
n_{\rm max} \approx  \dfrac{\lambda_0}{2 d_z},
}
and grows indefinitely with shrinking lattice constant. In reality, the band structure is not always invertible, due to the interfering mechanisms of energy transfer between planes via radiation and evanescent waves. For fixed $d_z/d$, non-invertibility will arise for sufficiently small $d$, while for fixed $d$, increasing $d_z/d$ will eventually lead to invertibility. This is illustrated in Fig. \ref{fig:fra_band_and_scaling}, for example, as the choices $d/\lambda_0=1/10$,~$d_z/d=1$ and $d/\lambda_0=1/60$,~$d_z/d=1.6$ are invertible, while $d/\lambda_0=1/60$,~$d_z/d=1$ is not. The condition for invertibility is derived in greater detail in Appendix \ref{app:invertible_band}. In what follows, we will fix $d_z/d=2.5$, where the contribution of the evanescent coupling to the dispersion relation is negligible $|J_{\text{ev}}(k_z)|\ll |J(k_z)|$ down to $d\approx \lambda_0/360$ (corresponding to $d\approx 6a_0$ for hydrogen atoms) and the band remains invertible, by which point quantum chemistry has already become significant.

When the band is non-invertible, the index is not well-defined. In particular, incident light can excite different values of $k_z$, and ``split'' into components propagating at different phase velocities. In the presence of additional dissipation, such as arising from quantum chemistry, light strongly favors exciting the lower value of $k_z$. This leads to a practically observed index that can be much smaller than the prediction of Eq.~(\ref{eq:fra_max_index_invertible}).

We conclude this subsection by noting that the idea of using resonant scatterers to potentially realize high-index materials has been discussed before, typically in the context of small metallic nanoparticles or metal composites with plasmonic resonances~\cite{khurgin_expanding_2022,shim_fundamental_2021}. Compared to such works, two key differences of our work are that first, we consider isolated atoms as building blocks that are completely lossless and have a large scattering cross section decoupled from their physical size, and that second, by bringing the atoms progressively closer until quantum chemistry turns on, we can better address the fundamental limits of refractive index of a ``real'' material. 

\subsection{Collective versus distinguishable response}
\label{subsec:distinguishable}

\begin{figure}[t]
\centering
\includegraphics[width=0.7\textwidth]{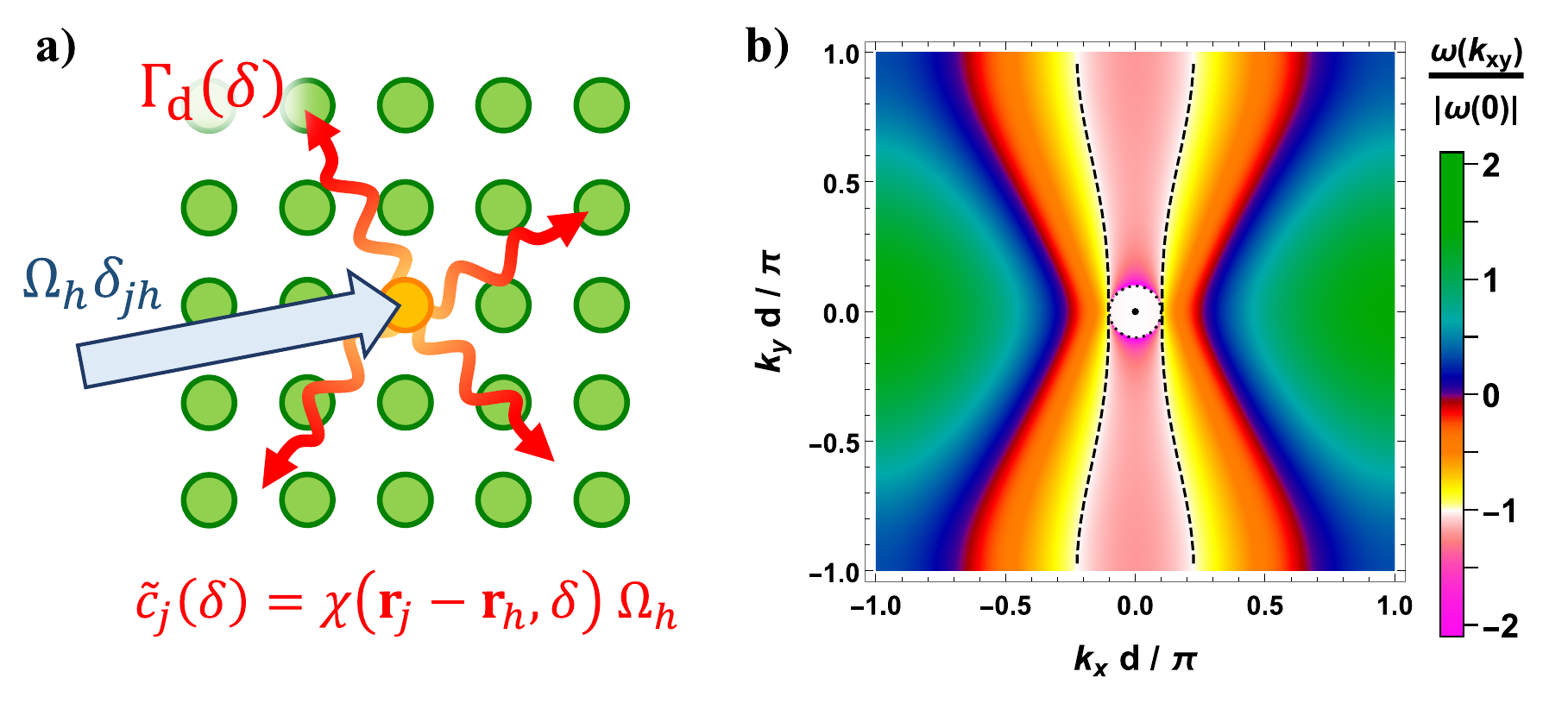}
\caption{a) A single, distinguishable atom at $\bfr_h$ in a 2D array is selectively driven by a weak input field of Rabi frequency $\Omega_h$~(blue arrow) and detuning $\delta$. Dipole-dipole interactions between atoms allow other atoms $j$ in the array to become excited by an amount proportional to the susceptibility $\chi(\bfr_j-\bfr_h,\delta)$. The rate of such a process is given by $\Gamma_{\rm d}(\delta)=-2\im[\Omega_h/\tilde c_h(\delta)]=-2\im 1/\chi(0,\delta)$. b) Dispersion relation $\omega(\bfk_{xy})$ of a 2D atomic array, within the first Brillouin zone. The lattice constant is chosen as $d=\lambda_0/20$, while the black, dashed line and the black point at the center represent the isoenergetic modes with $\omega(\bfk_{xy})=\omega(0)$. The modes with $|\bfk_{xy}|<k_0$ that radiate into free space are inside the boundary given by the dotted black circle.}
\label{fig:distinguishable}
\end{figure}

The collective response of a uniformly excited array differs remarkably from the response of a single, distinguishable driven atom. Concretely, we now consider an infinite 2D array, but where a weak input field with detuning $\delta$ selectively drives just a single atom located at $\bfr=\vecTemp r _{h}$, i.e. taking $\Omega_j=\Omega_{h} \delta_{jh}$ in Eq.~(\ref{eq:3DQO_Hamiltonian}), as illustrated in Fig.~\ref{fig:distinguishable}a. Although this scenario might not appear particularly physical, it allows us to derive a response function that will be directly relevant for our quantum chemistry discussions later. In particular, we will see in Sec.~\ref{sec:indexchemistry} that it characterizes the optical response of photo-excited electrons that tunnel to neighboring nuclei.

We consider a wave function $|\psi_{\rm 2D}\rangle=c_G|G\rangle+\sum_j \tilde{c}_j b_{pj}^{\dagger}b_{sj}|G\rangle$, where other atoms $j\neq h$ can still be excited via dipole-dipole interactions with the driven atom. Our goal is to solve for the steady-state atomic amplitudes $\tilde{c}_j(\delta)$ under Eq.~(\ref{eq:3DQO_Hamiltonian}), assuming that $c_G\approx 1$. This can be efficiently done by calculating the free propagator inside the 2D atomic array, which describes the spread of the excitation mediated by dipole-dipole interactions. 
In the rotating frame, such a propagator is given by the operator $G_{\chi}(\delta)=-(H_0+H_{\rm dip-dip})^{-1}$, which can be explicitly computed by decomposing the single-excitation manifold into the Bloch modes $|E_{{\bfk}_{xy}}\rangle$ that diagonalize it. One obtains
\eq{
\label{eq:free_propagator_G_chi}
G_{\chi}(\delta)= -N\left(\dfrac{d}{2\pi}\right)^2 \Int_{\text{BZ}} \text d\vecTemp k_{xy} \dfrac{\ket{E_{\vecTemp k_{xy}}}\bra{E_{\bfk_{xy}}}}{-\delta+\omega(\vecTemp k_{xy})-i \Gamma(\vecTemp k_{xy})/2} = -\sum_{jh\sigma\sigma'} \chi(\vecTemp r_j-\vecTemp r_h,\delta)\;(b_{pj\sigma}^{\dagger}b_{sj\sigma})(b^{\dagger}_{sh\sigma'}b_{ph\sigma'}),
}
where we have defined the susceptibility
\eq{
\label{eq:susceptibility_def}
\chi(\vecTemp r_j - \vecTemp r_h,\delta) = \left(\dfrac{d}{2\pi}\right)^2 \Int_{\text{BZ}} \text d\vecTemp k_{xy} \dfrac{e^{- i \vecTemp k_{xy}\cdot (\vecTemp r_j - \vecTemp r_h)}}{-\delta+\omega(\vecTemp k_{xy})-i \Gamma(\vecTemp k_{xy})/2}.
}
The physical meaning of these operators can be seen by examining the first-order expansion of the Dyson equation, which leads to the weak-driving steady state 
$\ket{\psi_{\rm 2D}}=(1+G_{\chi}(\delta)H_{\rm drive})\ket{G}$. For the case of a selectively driven atom, one has $H_{\rm drive}=-\Omega_h \sum_{\sigma}\left( b^{\dagger}_{ph\sigma}b_{sh\sigma}+h.c.\right)$, which leads to the coefficients
\begin{equation}
\tilde{c}_j(\delta) = \chi(\vecTemp r_j-\vecTemp r_{ h},\delta) \;\Omega_{ h}.\label{eq:tildech}
\end{equation}
%


Some of the energy provided by the drive will naturally be radiated into free space, through the excitation of collective modes $|\bfk_{xy}|\leq k_0$ with non-zero radiative decay rate $\Gamma(\bfk_{xy})$. However, at small lattice constants $d\ll \lambda_0$ this channel is negligible compared to the amount of energy that has gone into exciting non-radiative modes with $|\bfk_{xy}|>k_0$, which subsequently propagate outward from $\bfr_h$ along the array itself. To illustrate this, we first plot $\omega(\bfk_{xy})$ within the first Brillouin zone $|k_x|,|k_y|<\pi/d$ in Fig. \ref{fig:distinguishable}b for a lattice constant $d=\lambda_0/20$ that is small compared to the resonant wavelength. For $d\ll\lambda_0$, the first Brillouin zone is dominated by the region outside the light cone $|\bfk_{xy}|>k_0$. 
Restricting the integration in \eqrefTemp{eq:susceptibility_def} to this dominant region, one has $\Gamma(\vecTemp k_{xy})=0 $, while the energy scale is dictated by the near-field ($\sim 1/r^3$) component of the Green's function, leading to the functional form $\omega(\bfk_{xy})\sim \Gamma_0(\lambda_0/d)^3 f(\bfk_{xy}d/\pi)$. Considering that the region of integration scales as $\text d\vecTemp k_{xy}= (\pi/d)^2 \text d(\bfk_{xy}d/\pi)$, one then obtains the final scaling $ \chi (\vecTemp r_j-\vecTemp r_h,\delta)\sim d^3/(\Gamma_0 \lambda_0^3)$. We note that $\chi(0,\delta)$ can have an imaginary component describing work done by the drive on atom $h$. This occurs if there exists an isoenergy contour where $\omega(\bfk_{xy})=\delta$~(see the dashed black curve in Fig.~\ref{fig:distinguishable}b for the contour $\omega(\bfk_{xy})=\omega(0)$), allowing the drive to resonantly excite a continuum of non-radiative modes. Specifically, the quantity $\Gamma_{\rm d}(\delta)=-2\im[ \Omega_h/\tilde c_h(\delta)] = -2\im 1/\chi(0,\delta)$ quantifies the rate at which energy is irradiated into the atomic array via the selectively driven atom (as pictorially described by the wavy arrows in \figref{fig:distinguishable}a). In Appendix~\ref{app:distinguishable_atom_gamma_d}, we describe our numerical procedure to calculate $\chi$, which demonstrates the scaling mentioned above.

For the quantum chemistry problem, it will also be helpful to understand the problem of a 2D array illuminated by a normally incident plane wave of Rabi frequency $\Omega_0$, but with a single missing atom at site $\bfr=\bfr_h$. In Sec.~\ref{sec:indexchemistry}, we will see that this classical calculation roughly equates with the optical response arising from electronic density-density correlations. Working in the usual weak driving limit, it is convenient to write the steady-state, single-excitation amplitude of atom $j$ as $c_j(\delta)=c_E(\delta)+\tilde{c}_j(\delta)$. Here, $c_E(\delta)$ is the solution for a defect-free, uniformly driven array given in Eq.~(\ref{eq:cE}), while $\tilde{c}_j(\delta)$ is the solution for an array with a single driven atom at $\bfr_h$ given in Eq.~(\ref{eq:tildech}). This expression is valid provided that $\Omega_h=-c_E(\delta)/\chi(0,\delta)$ is chosen such that $c_h(\delta)=c_E(\delta)+\tilde{c}_h(\delta)=0$. Physically, this states that the overall solution can be expressed as the coherent sum of the solution of two separate problems, of a uniformly driven perfect array and a perfect array with a single driven atom. Enforcing that $c_h(\delta)=0$ via a proper choice of the single-atom driving amplitude $\Omega_h$ says that the atom at $\bfr_h$ has no excitation amplitude, which is equivalent to having no atom at $\bfr_h$ to begin with.

To quantify the effect of a missing atom, we can calculate the scattering cross-section associated with a hole at the driving frequency $\delta = \omega(0)$. In general, given a set of atomic dipoles illuminated by a field $\Omega_j$ and producing coefficients $c_j$, the total cross section is given by the optical theorem~\cite{newton_optical_1976,alaee_kerker_2020} to be
$
\sigma=(\sigma_0/2)(\Gamma_0/\Omega_0^2)\im \sum_j  \Omega_j c_j ,
$
where $\sigma_0=3\lambda_0^2/(2\pi)$ is the resonant cross section of a single atom in vacuum. It is convenient to normalize this by the cross-section of a single atom in the perfect array at the same frequency. Defining this ratio as $N_h$, we find that 
%
\begin{equation}
N_h = \dfrac{  \im \Omega_h\tilde{c}_h  }{\im \Omega_0 c_E } = \left. \dfrac{\Gamma_{\rm d}(\delta=\omega(0))}{\Gamma(0)}\right.\sim \dfrac{\lambda_0}{d}.\label{eq:Nh}
\end{equation}
$N_h$ can be interpreted as the effective number of atoms affected by the single-site hole, or equivalently the size of the ``effective'' hole as seen by resonant, incident light. In particular, due to the $\lambda_0/d$ scaling, for small $d$ the effective hole size can be much bigger than the unit cell size $d^2$, and $N_h\gg 1$.

The total field produced by a plane wave interacting with an array with a single hole can be derived from Eq.~(\ref{eq:input-output_full}) and the excitation amplitudes $c_j(\delta)$. We can also generalize to the case of an array with a small fraction $P_h\ll 1$ of defects randomly removed, if we assume that the defects are sufficiently far enough apart that their emission is uncorrelated. This results in a generalized transmission coefficient of (compare with Eq.~(\ref{eq:diffracted_field_subwavelength})) of 
\eq{
\label{eq:t_holes}
t(\delta)
\approx 1+\dfrac{i\Gamma(0)/2}{-\delta +\omega(0) -i \Gamma(0)  /2 + P_{\text h} \Sigma _{\text{h}}(\delta)},
}
and a reflection coefficient $r(\delta)=t(\delta)-1$. We see that the effect of the holes can be incorporated into a complex ``self-energy'' $\Sigma_{\text{h}}(\delta)=1/\chi(0,\delta)$, describing an effective energy shift and additional decay rate experienced by the collective mode $|E_{\bfk_{xy}=0}\rangle$. In particular, this additional decay implies that $|r(\delta)|^2+|t(\delta)|^2<1$, due to the holes scattering light into other directions $\bfk_{xy}\neq 0$.

\section{Quantum chemistry model}
\label{sec:model}

\subsection{Introduction to model}

The potentially large and purely real refractive index obtained in the quantum optics limit, $n_{\rm max}\approx \lambda_0/2d_z$, is associated with the single-mode nature of the light-matter interaction. Intuitively, the new degrees of freedom and dynamics opened up by quantum chemistry can break the single-mode nature, creating channels for inelastic or spatial multimode scattering and subsequently reducing the maximum index.

As mentioned in the introduction, our primary goal here is to understand the behavior of the refractive index at densities corresponding to the onset of quantum chemistry, when the lattice constant is still large compared to the Bohr radius, $d\gg a_0$. We favor this approach because chemistry can be considered weak and can thus be treated perturbatively, which allows one to avoid the well-known theoretical and computational challenges of quantum chemistry of solids. Studying this regime also enables one to continue to treat multiple scattering non-perturbatively, which is key to understanding the limits of refractive index.

Within the weak chemistry limit, one still has to choose the individual atomic building block of the lattice. We take hydrogen atoms, which have the advantage that the single hydrogen atom is an exactly solvable quantum mechanics problem. One could instead conceivably take an atom that corresponds to a more realistic optical solid~(e.g., silicon), with the price that the single atom is already a complicated many-electron problem, and density functional theory or other techniques would have to be applied to justifiably and quantitatively reduce to a more minimal model~(such as only involving the valence electrons). Despite the specificity of taking hydrogen, we will see that the main mechanisms that limit the refractive index involve the emergence of quantum magnetism, chemistry-induced electronic density-density correlations, and tunneling dynamics of photo-excited electrons. These are rather general features in materials, which plausibly give our model broader qualitative validity. Finally, in our model, we assume that the nuclei can be magically ``fixed'' to realize any lattice constant,  while more realistic routes toward an ultrahigh index material are discussed in Sec.~\ref{subsec:index_limit}. 

To be specific, we consider a rectangular lattice with lattice constant $d$ in the transverse plane and $d_z=2.5d$ along the direction of light propagation, avoiding the non-invertible optical band structure discussed in Sec.~\ref{subsec:indexQO}. For an isolated hydrogen atom, the transition wavelength from the 1s to 2p level is $\lambda_0\approx 121$~nm and the corresponding spontaneous emission rate is $\Gamma_0\approx 2\pi\times 100$~MHz. Of course, neither hydrogen nor any other material is energetically stable for arbitrary values of $d,d_z$, but again we assume that the nuclei can be fixed for this thought experiment.

Formally, the Hamiltonian describing the quantum chemistry and the light-matter interactions associated with the hydrogen lattice is given by %
\begin{equation}
    H=H_{\rm el} + H_{\rm el-el}+H_{\rm ph}+H_{\rm ph-el}+H_{\rm drive}\,. \label{eq:H}
\end{equation}
Here, $H_{\rm el}=\sum_i h_i$ where
\begin{equation}
h_i=\frac{\bfp_i^2}{2m}-\sum_j V_C(\bfr_i-\bfR_j) \end{equation}
describes the kinetic energy of electron $i$ and its Coulomb interaction  $V_C(\bfr)=q^2/4\pi\epsilon_0|\bfr|$ with the positive nuclei fixed at positions $\bfR_j$. The sums run over $1\leq i,j \leq N$, where for hydrogen $N$ corresponds both to the number of nuclear sites and the number of electrons in the system. The second term in Eq.~\eqref{eq:H} captures the electrostatic interaction between the electrons and takes the form $H_{\rm el-el}=(1/2)\sum_{i,j\neq i}V_C(\bfr_i-\bfr_j)$. The third term describes free photons $H_{\rm ph}=\sum_{\bfk} \hbar\omega_{\bfk} a^{\dagger}_{\bfk}a_{\bfk}$ with dispersion relation $\omega_{\bfk}=c|\bfk|$, and the fourth their interaction with matter, which in the Coulomb gauge reads
\begin{equation}
    H_{\rm ph-el}=-\frac{q}{m}\sum_i\vecTemp{A}(\bfr_i)\cdot\bfp_i+\frac{q^2}{2m}\sum_i\vecTemp{A}(\bfr_i)^2. \label{eq:H_ph}
\end{equation}
Here, $\vecTemp{A}(\bfr)$ denotes the vector potential of the electromagnetic field which admits the mode decomposition
\begin{equation}
    \vecTemp{A}(\bfr)=\sum_{\bfk\alpha}\hat{\mathbf{e}}_\bfk^\alpha\sqrt{\frac{\hbar}{2\omega_\bfk\varepsilon_0V}}\left(a_\bfk e^{i\bfk\cdot\bfr}+h.c.\right)\,, \label{eq:vector_potential}
\end{equation}
where $V$ is the quantization volume, and $\hat{\mathbf{e}}_\bfk^\alpha$ with $\alpha\in\{1,2\}$ denote the transverse polarization unit vectors obeying $\hat{\mathbf{e}}_\bfk^\alpha\cdot\bfk=0$. The final term in Eq.~\eqref{eq:H} is associated with the (classical) optical driving field.

We bring the matter part of the Hamiltonian into second quantized form by associating a localized, orthonormal set of electronic Wannier states $|\phi_{i\nu\sigma}\rangle$ centered around each site $i$, where $\sigma$ denotes the spin state and $\nu$ the band index. The associated fermionic creation operators $b_{\nu i\sigma}^{\dagger}$ were already introduced in Eq.~\eqref{eq:3DQO_Hamiltonian}. The Wannier functions at different sites are related by translational symmetry,  $|\phi_{i\nu\sigma}\rangle=|\phi_\nu(\bfr-\bfR_i)\rangle|\sigma\rangle$, where the notation $|\phi_\nu(\bfr-\bfR_i)\rangle$ indicates that the projection of this state onto a position basis yields a wavefunction $\phi_\nu(\bfr-\bfR_i)$ with $\phi_\nu(\bfr)$ exponentially localized in $\bfr$. In the non-interacting limit of large $d/a_0\rightarrow \infty$, the form of the Wannier orbitals $\phi_\nu(\bfr)$ approaches that of the atomic orbitals. In terms of the Wannier states and operators,
\begin{eqnarray}
    H_{\rm el} & = & \sum_{ij,\nu,\sigma} \langle\phi_{\nu}(\bfr-\bfR_i)|h|\phi_{\nu}(\bfr-\bfR_j)\rangle b_{\nu i \sigma}^{\dagger}b_{\nu j \sigma} \nonumber \\ 
    H_{\rm el-el} & = & \frac{1}{2}\sum_{ijkl,\mu\mu'\nu\nu',\sigma\sigma'}\langle\phi_{\mu}(\bfr-\bfR_i)\phi_{\mu'}(\bfr'-\bfR_j)|V_C(\bfr-\bfr')|\phi_{\nu'}(\bfr'-\bfR_k)\phi_{\nu}(\bfr-\bfR_l)\rangle b_{\mu i \sigma}^{\dagger} b_{\mu' j \sigma'}^{\dagger} b_{\nu' k \sigma'} b_{\nu l \sigma} \nonumber \\
    H_{\rm ph-el} & = & \sum_{ij,\nu\nu',\sigma} \langle\phi_{\nu}(\bfr-\bfR_i)|\left(\frac{q^2}{2m}\vecTemp{A}(\bfr)^2-\frac{q}{m}\vecTemp{A}(\bfr)\cdot\bfp\right)|\phi_{\nu'}(\bfr-\bfR_j)\rangle\,b^{\dagger}_{\nu i \sigma}b_{\nu'j\sigma}.
    \label{eq:Wannier_expansion}
\end{eqnarray}
Above, we note that $H_{\rm el}$ is block diagonal in the band index $\nu$.

\subsection{Effective 2D Hamiltonian}\label{subsec:2DHamiltonian}

\subsubsection{Simplifying assumptions}\label{subsubsec:simplifying}
While the Hamiltonian of Eq.~(\ref{eq:H}) is completely general, we now introduce simplifications we can make in the limit $d_z=2.5d$ and $d\gg a_0$:
\begin{itemize}
    \item For these lattice constants, we can assume that quantum chemistry first turns on within individual 2D layers, while chemical interactions between layers~(separated by $d_z$) remain negligible.
    \item Being interested in near-resonant light interactions, we take the long-wavelength limit. Physically, we assume that the field interacting with each atom does not probe the spatial extent of the associated electronic orbital around the nucleus, which implies the approximate eigenvalue equation $\bfr\ket{\phi_\nu(\bfr-\bfR_i)}\approx\bfR_i\ket{\phi_\nu(\bfr-\bfR_i)}$ as far as light is concerned. In particular, the interaction term $H_{\rm ph-el}$ in Eq.~\eqref{eq:Wannier_expansion} takes the approximate form
    \begin{equation}
        H_{\rm ph-el}\approx-\frac{q}{m}\sum_{ij,\nu\nu',\sigma}\vecTemp{A}(\bfR_i)\cdot\langle\phi_{\nu}(\bfr-\bfR_i)|\bfp|\phi_{\nu'}(\bfr-\bfR_j)\rangle\,b^{\dagger}_{\nu i \sigma}b_{\nu'j\sigma}\label{eq:H_longwavelength}
    \end{equation}
    where we have dropped the term quadratic in $\vecTemp A(\bfR_i)$ that no longer couples to the electronic states. We further work in the regime of weak driving, which is sufficient to probe the linear refractive index, and restrict our calculations to the two lowest electronic bands (those reducing to the 1s and 2p hydrogen levels in the isolated atom limit).
    \item In the large $d/a_0$ limit, the Wannier orbitals have an exponentially reduced weight at neighboring nuclei. We use this to truncate Eq.~(\ref{eq:Wannier_expansion}) to the following terms: on-site terms and tunneling between nearest neighbor nuclei in $H_{\rm el}$, on-site $H_{\rm el-el}^{(0)}$~(with $i=j=k=l$) and pair~($i=l, j=k$) interactions in $H_{\rm el-el}$, and on-site terms~($i=j$) in $H_{\rm ph-el}$.
    \item The on-site terms captured by $H_{\rm el-el}^{(0)}$ cause states where two electrons sit on the same nucleus to have a large energy cost $U$, which is on the order of the hydrogen ionization energy. This large energy cost allows the dynamics to be projected into an effective low-energy Hamiltonian, within the manifold of one electron per nucleus.
    \item We integrate out the photons, which along with the pair interaction terms in $H_{\rm el-el}$, give rise to the dipole-dipole interactions previously analyzed in Sec.~\ref{sec:QO}.
    \end{itemize}

\subsubsection{Presentation of effective 2D Hamiltonian}\label{subsubsec:presentation}
Under these conditions, the effective Hamiltonian governing atom-light interactions and quantum chemistry of a 2D array is given approximately by
\begin{eqnarray}
    H_{\rm 2D} & = & \underbrace{\sum_{i\sigma}\left(\epsilon_s b_{si\sigma}^{\dagger}b_{si\sigma}+\epsilon_p b_{pi\sigma}^{\dagger}b_{pi\sigma}\right)}_{H_0}  \underbrace{-\Gamma_0\sum_{ij,\sigma\sigma'}G_{ij}\left(b_{pi\sigma}^{\dagger}b_{si\sigma}\right)\left(b_{sj\sigma'}^{\dagger}b_{pj\sigma'}\right)}_{H_{\rm dip-dip}} + \underbrace{\sum_{i\sigma}\left(\Omega e^{-i\omega_L t}b_{pi\sigma}^{\dagger}b_{si\sigma}+h.c.\right)}_{H_{\rm drive}} \nonumber \\ & & + \underbrace{J\sum_{\langle ij\rangle} \bm{S}_i\cdot \bm{S}_j - t_{\rm eff}\sum_{\langle ij\rangle \sigma\sigma'} (b_{pi\sigma}^{\dagger}b_{pj\sigma}b_{sj\sigma'}^{\dagger}b_{si\sigma'}+h.c.)}_{H_{\rm tJ}=H_J+H_t} \nonumber \\ & & = H_{\rm QO}+H_{\rm tJ} \label{eq:H2D}
\end{eqnarray}
where $\langle ij \rangle$ denotes restriction of the sum to nearest neighbors, while $\bm{S}_i=(1/2)\sum_{\sigma\sigma'}b_{si\sigma}^{\dagger}\vec{\tau}_{\sigma\sigma'}b_{si\sigma'}$ and $\vec{\tau}$ denotes the Pauli matrices. Naturally, one can see that a subset of the terms in $H_{\rm 2D}$~(the first three terms on the right hand side) correspond to the Hamiltonian $H_{\rm QO}$, Eq.~(\ref{eq:3DQO_Hamiltonian}), in the quantum optics limit, here written without the rotating frame and implicitly restricted to a 2D array of atoms. We note that while the assumptions stated in Sec.~\ref{subsubsec:simplifying} leading to the above Hamiltonian are reasonable, it is difficult to completely quantify the errors associated with the terms that are omitted with respect to the full Hamiltonian~(\ref{eq:H}). Indeed, if that could be done, that would amount to doing exact quantum chemistry, which is likely intractable. Eq.~(\ref{eq:H2D}) should thus be viewed as a minimal model that is believed to contain the key physics. We now provide some intuition of the physics described by the ``new'' part of the Hamiltonian $H_{\rm tJ}$ arising from quantum chemistry. A more detailed derivation of Eq.~(\ref{eq:H2D}) is given in Sec.~\ref{sec:derivation} for those readers who are interested.

\begin{figure}[t]
\centering
\includegraphics[width=0.85\textwidth]{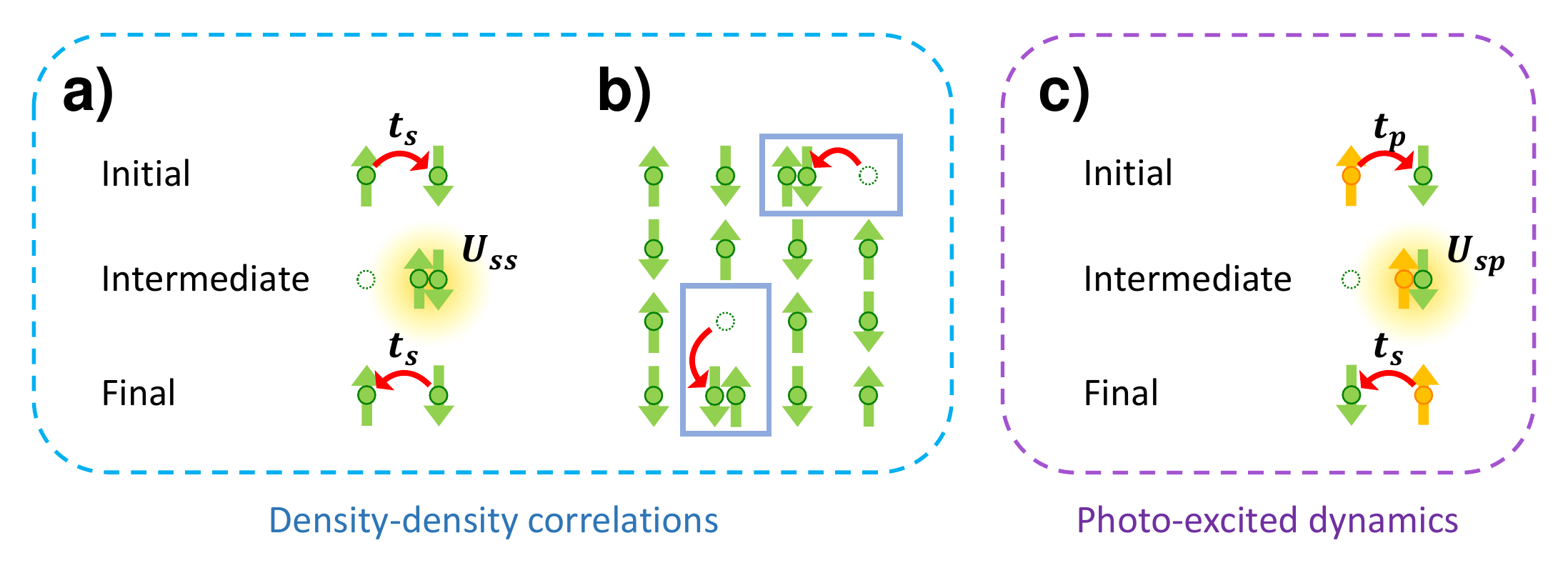}
\caption{a) Illustration of the perturbative process giving rise to anti-ferromagnetic Heisenberg spin interactions between nearest-neighbor electrons in their s-orbitals. An electron tunnels to its nearest neighbor, giving rise to an intermediate state with large energy $E=U_{ss}$ due to on-site interactions, and an electron subsequently tunnels back. This process is only allowed provided the two electrons have different spins, due to Pauli exclusion. b) The Heisenberg spin interaction leads to an anti-ferromagnetic N\'{e}el order for the many-body ground state, as qualitatively illustrated here by the checkerboard pattern of up and down spins of the s-orbital electrons. At next order in perturbation theory, the intermediate state illustrated in a) manifests itself in the many-body ground state through the appearance of bound holon-doublon pairs~(the pairs of sites outlined by rectangles), which reflect the electronic density-density correlations generated by the interactions. These pairs consist of a holon~(dashed circle), i.e. a nucleus without an electron, and a neighboring doublon with two electrons. c) An analogous process to a) can occur if an electron on one site is in its p-orbital~(with the p-orbital state indicated by orange) and a neighbor is in its s-orbital~(indicated by green). Note that going from the initial to the final state, both the spin and orbital degrees of freedom between the sites have been exchanged.
}
\label{fig:2Dexplained}
\end{figure}

The Hamiltonian $H_{\rm tJ}$ coincides with the tJ model Hamiltonian of condensed matter physics. In particular, the on-site interaction energy $U_{ss}$ for two s-orbital electrons to occupy the same nucleus is on the order of the hydrogen ionization energy, while the tunneling rate $t_s$ for such electrons is exponentially suppressed for large $d/a_0$. Thus, one has that $U_{ss}\gg |t_s|$, in which case the site occupancy of the ground state is frozen to one s-orbital electron per site~(Mott insulator). However, within second-order perturbation theory, an electron can tunnel to its nearest neighbor and back, provided that the involved electrons are of opposite spin, as illustrated in Fig.~\ref{fig:2Dexplained}a. This gives rise to the anti-ferromagnetic Heisenberg interactions $H_J$ between the spin degrees of freedom of nearest neighbor electrons. This reflects the onset of quantum magnetism due to quantum chemistry, and is well-known to arise from the single-band, half-filled Fermi-Hubbard model, in the limit of $U_{ss}\gg |t_s|$~\cite{auerbach_interacting_1994}. The spin interaction strength is given by $J=4t_s^2/U_{ss}$.

As a result, the global state $|\sigma\rangle$ of the spins in the ground state has anti-ferromagnetic N\'{e}el order, as qualitatively illustrated in Fig.~\ref{fig:2Dexplained}b. This alone does not alter the optical properties discussed in Sec.~\ref{sec:QO}, as the spin is decoupled from electron-photon interactions. As far as ground state properties are concerned, the first chemical effect that modifies the optical response comes from considering the change in total wave function at the next order of perturbation theory. Due to its complexity, we do not explicitly include it in $H_{\rm tJ}$. However, it is qualitatively easy to understand and is discussed in more detail in Sec.~\ref{sec:derivation}.

Specifically, while $H_J$ describes the perturbative effect of tunneling within the low-energy manifold of one electron per nucleus, the intermediate state in Fig.~\ref{fig:2Dexplained}a at next order of perturbation theory leads to a total ground state illustrated in Fig.~\ref{fig:2Dexplained}b, where the number of electrons per site is no longer fixed to one, and there is a small probability $\sim (t_s/U_{ss})^2$ to find holon-doublon pairs consisting of two electrons on one nucleus and no electrons on a nearest neighbor. A more precise many-body calculation presented in Sec.~\ref{sec:derivation} shows that the fraction of sites occupied by holons or doublons is $2P_{\rm hd}\approx 5.16(t_s/U_{ss})^2$, where $P_{\rm hd}$ indicates the ratio of the number of holon-doublon pairs over the total number of lattice sites. These holon-doublon pairs are a manifestation of electronic density-density correlations that emerge due to quantum chemistry, and we will later argue that light sees these pairs as effective holes that reduce the optical response of the 2D array.

Having described the relevant physics of the many-body ground state, we now turn to the dynamics that can occur upon excitation with weak light, when an electron will be promoted to a p-orbital. The dynamics of the photo-excited electron is contained in the Hamiltonian term $H_t$. Physically, it describes motion of an excited p-orbital among the background of s-orbital electrons through the perturbative process illustrated in Fig.~\ref{fig:2Dexplained}c. For example, a p-orbital electron can first tunnel to a nearest neighbor already containing an s-orbital electron, to create a high-energy intermediate state of energy $E=U_{sp}$. The s-orbital can then tunnel back to replace the p-orbital on its original site. Within perturbation theory, the overall rate of such processes is given by $t_{\rm eff}=2t_s t_p/U_{sp}$, where $t_p$ is the tunneling amplitude of the p-orbital electron, and $U_{sp}$ is the on-site energy associated with having an s- and p-orbital sitting on the same nucleus. Note that this double tunneling event not only exchanges the p-orbital and s-orbital electrons at neighboring sites, but also their spin states. The s-orbital electron spins that are displaced in this way by the excited electron dynamics break the original anti-ferromagnetic order and thus incur a spin interaction energy cost. As discussed further in Sec.~\ref{sec:indexchemistry}, this energy loss due to spin flips results in inelastic photon emission when the p-orbital drops back down to an s-orbital, and presents a limitation to the maximum achievable refractive index.

The Fermi-Hubbard model or extensions of it have served as the starting point for various studies of the dynamics of photo-excited electrons~(see, e.g., Ref.~\cite{eckstein_ultrafast_2014}). Compared to previous work, a key difference in our work is the inclusion of an additional, long-range dipole-dipole interaction $H_{\rm dip-dip}$ in Eq.~(\ref{eq:H2D}), which encodes the possibility of non-perturbative multiple scattering of light. From Sec.~\ref{sec:QO}, we see that this term is needed to correctly predict the large refractive index in the quantum optics limit, and apparently its reduction in the quantum chemistry regime.

\subsection{Energy scales of interactions}\label{subsec:scales}

Here, we quantify how the various energy scales behave as a function of $d/a_0$. In principle, the Wannier functions for a lattice of hydrogen ions could be numerically computed by standard techniques~\cite{marzari_maximally_2012} and the microscopic parameters $t_{s,p},U_{ss,sp}$ directly obtained. However, this is challenging in our regime of interest where $d/a_0\gg 1$, as one expects $t_{s,p}$ to be exponentially small. We therefore adopt an alternative strategy, assuming that the microscopic parameters $t_{s,p}$ also characterize the spectrum of the $\textrm{H}_2^+$ hydrogen molecule ion, and benefitting from the fact that the energy curves of this molecule can be calculated with very high numerical precision. We specifically utilize the numerical data of Ref.~\cite{fernandez_highly_2021}.

In particular, we equate $2t_s$ with the energy splitting between the two states of the ground state manifold~(with the labels  $\sigma_u^{\ast}1s$ and $\sigma_g1s$ in the conventional separated-atoms description). These two states approximately correspond to the odd/even superpositions of a 1s orbital on the two nuclei $a$ and $b$, $(1s)_a\mp (1s)_b$ at large nuclear separation. Likewise, we equate $2t_p$ with the splitting between the $\sigma_u^{\ast}2p$ and $\sigma_g 2p$ excited states, which roughly correspond to the states $(2p_x)_a\mp (2p_x)_b$ at large nuclear separation~(we take the axis of separation to correspond to $x$). Finally, we approximate the on-site energies at large $d/a_0$ from the orbital wave functions $\phi_{1s},\phi_{2p_x}$ of the isolated hydrogen atom, with
\begin{eqnarray} U_{ss} & = & \frac{q^2}{4\pi\epsilon_0} \int d\bfr_1 d\bfr_2 \frac{1}{|\bfr_1-\bfr_2|} \phi_{1s}^2(\bfr_1) \phi_{1s}^2(\bfr_2), \nonumber \\ U_{sp} & = & \frac{q^2}{4\pi\epsilon_0} \int d\bfr_1 d\bfr_2 \frac{1}{|\bfr_1-\bfr_2|} \phi_{1s}^2(\bfr_1) \phi_{2p_x}^2(\bfr_2).\label{eq:UssUsp} \end{eqnarray}
These energies can be evaluated exactly from the known wave functions of the hydrogen atom. As detailed in Sec.~\ref{subsec:tJ}, one finds that $U_{ss}=5|\epsilon_s|/4$ and $U_{sp}=118|\epsilon_s|/243$, where $\epsilon_s\approx -13.6$~eV is the hydrogen ground state energy. 

\begin{figure}[t]
\centering
\includegraphics[width=0.9\textwidth]{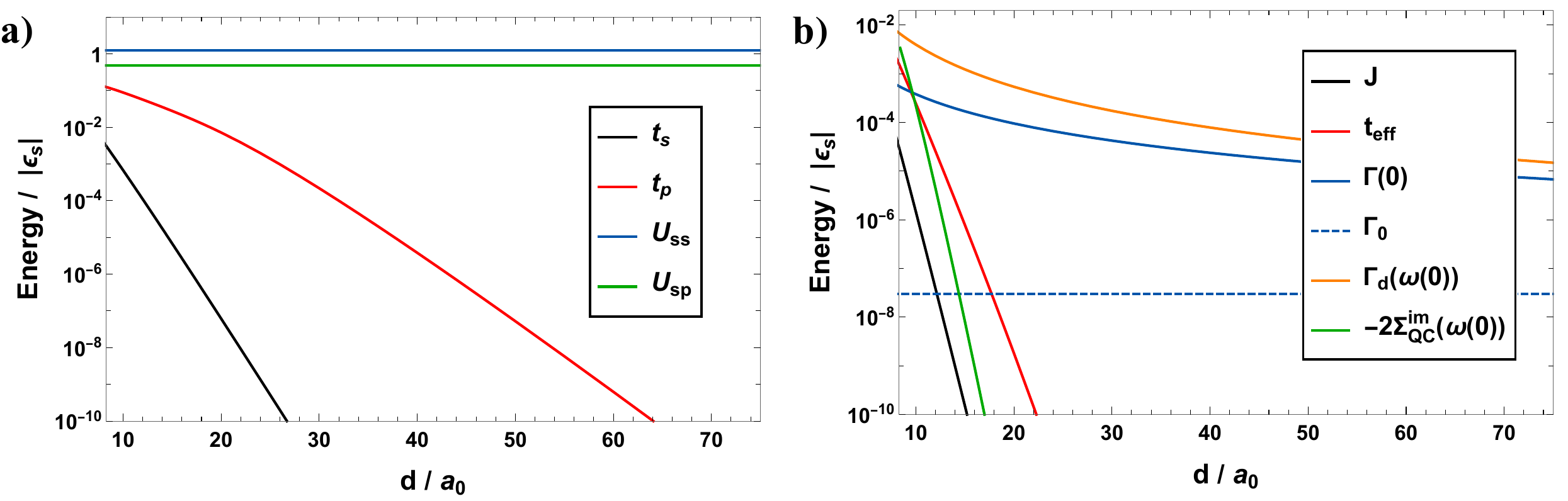}
\caption{a) Numerical values used in our model for the tunneling rates $t_s$~(black) and $t_p$~(red) of s- and p-orbital electrons, respectively, as a function of scaled lattice constant $d/a_0$. These rates are inferred from the spectrum of the $\textrm{H}_2^+$ hydrogen molecule ion. We also plot the on-site interaction energies $U_{ss}$~(blue) and $U_{sp}$~(green), as calculated from the isolated hydrogen atom orbitals. Energies are shown in units of the Rydberg constant $|\epsilon_s|$. b) Heisenberg interaction strength $J$~(black) and p-orbital impurity tunneling rate $t_{\rm eff}$~(red) as a function of lattice constant, and in units of the Rydberg constant. The solid and dashed blue curves indicate the collective emission rate $\Gamma(0)$ of an ideal 2D lattice in the quantum optics limit, and the emission rate $\Gamma_0$ of an isolated hydrogen atom, respectively. The orange curve represents the resonant emission rate 
$\Gamma_{\rm d}(\delta=\omega(0))$ 
related to a selectively driven atom in an array, as discussed in Sec. \ref{subsec:distinguishable}. The green curve captures the effective dissipation rate in the optical response of a 2D array, due to quantum chemistry. This is quantified by $-2\im \Sigma_{\rm QC}(\delta=\omega(0))$, where here we evaluate the self-energy on resonance.}
\label{fig:parameter_estimates}
\end{figure}

The resulting tunneling and on-site interaction energies are plotted as a function of lattice constant~(i.e. internuclear separation) in Fig.~\ref{fig:parameter_estimates}a, with energies in units of the Rydberg constant. In Fig.~\ref{fig:parameter_estimates}b, we plot the quantities $t_{\rm eff}$ and $J$ derived from these microscopic parameters. We also plot the collective decay rate $\Gamma(0)$ of a 2D array. Once the inequality $|t_{s,p}|\ll U_{ss},U_{sp}$ is no longer satisfied, corresponding to $d/a_0\lesssim 10$, one expects that our minimal model~(\ref{eq:H2D}) will break down~(see, e.g., Ref.~\cite{simkovic_extended_2020} for a study of the single-band Fermi-Hubbard model at half filling).

\section{Derivation of effective 2D Hamiltonian}\label{sec:derivation}

In this section, we derive in greater detail the various terms that appear in the Hamiltonian of Eq.~(\ref{eq:H2D}). None of the results here are original, but are instead included for completeness. Readers who want to directly understand the consequences of Eq.~(\ref{eq:H2D}) on the maximum refractive index can skip to Sec.~\ref{sec:indexchemistry}.

\subsection{tJ Hamiltonian}\label{subsec:tJ}
First, we derive the low-energy Hamiltonian $H_{\rm tJ}$ of a 2D array by integrating out high-energy states associated with double occupation of a nuclear site. Starting from the full Hamiltonian of Eq.~(\ref{eq:H}), and taking the limits of interest discussed in Sec.~\ref{subsec:2DHamiltonian}, we expand the single-electron Hamiltonian $H_{\rm el}=H_{\rm el}^{(0)}+H_{\rm el}^{(1)}$ in terms of on-site and nearest neighbor terms, and consider only the on-site terms in the electron-electron interaction $H_{\rm el-el}^{(0)}$ that are energy conserving~(i.e. conserve the occupation number of the s- and p-orbitals individually). Written out, one has
\begin{equation}
    H_{\rm el-el}^{(0)}=\sum_{i}U_{ss}n_{si\uparrow}n_{si\downarrow}+U_{sp}\left(n_{pi\uparrow}+n_{pi\downarrow}\right)\left(n_{si\uparrow}+n_{si\downarrow}\right)+\sum_{i\sigma\sigma'}U_{sp,sp}b_{si\sigma}^{\dagger}b_{pi\sigma'}^{\dagger}b_{si\sigma'}b_{pi\sigma}.
\end{equation}
Here, $U_{ss}$ and $U_{sp}$ are defined as in Eq.~(\ref{eq:UssUsp}), while
\begin{equation}
    U_{sp,sp}= \frac{q^2}{4\pi\epsilon_0} \int d\bfr_1 d\bfr_2 \frac{1}{|\bfr_1-\bfr_2|} \phi_{1s}(\bfr_1)\phi_{2p_z}(\bfr_2)\phi_{1s}(\bfr_2)\phi_{2p_z}(\bfr_1).
\end{equation}

Note that the interaction energy arising from $U_{sp}$ is independent of spin state, while that from $U_{sp,sp}$ is spin dependent. These interaction energies can be evaluated exactly by taking the known hydrogen atom orbital wave functions, and by writing the interaction potential in terms of spherical harmonic functions, $|\bfr_1-\bfr_2|^{-1}=\sum_{l=0}^{\infty}\sum_{m=-l}^{l}\frac{4\pi r_{<}^l}{(2l+1)r_{>}^{l+1}}Y_{lm}^{\ast}(\theta_1,\phi_1)Y_{lm}(\theta_2,\phi_2)$. Here, $r_{<}$ and $r_{>}$ denote the minimum and maximum of $r_1,r_2$, respectively. As noted earlier, one finds $U_{ss}=5|\epsilon_s|/4$ and $U_{sp}=118|\epsilon_s|/243$, while evaluation of the spin-independent energy gives $U_{sp,sp}\approx 0.07 U_{sp}$. This allows us to ignore the small spin-dependent term $U_{sp,sp}$ from here on. The primary effect of the long-range nature of the Coulomb interaction~(beyond on-site terms of $H_{\rm el-el}^{(0)}$) is a contribution to dipole-dipole interactions, analyzed further in Section~\ref{subsec:electrostatic}. In the large $d/a_0$ limit, the remaining energy non-conserving on-site terms~(along with contributions from higher bands) would in principle provide an exact treatment of the negative hydrogen ion~(single proton, two electrons)~\cite{lin_ground_1975,hill_proof_1977}, but should not qualitatively change the subsequent results.

Written out explicitly, the subset of terms we analyze here is then given by
\begin{equation}
H' = \underbrace{\sum_{i\nu\sigma} \epsilon_\nu n_{\nu i \sigma}-\sum_{\langle ij \rangle\nu\sigma} t_\nu \left(b_{\nu i \sigma}^{\dagger}b_{\nu j\sigma}+h.c.\right)}_{H_{\rm el}^{(0)}+H_{\rm el}^{(1)}}+\underbrace{\sum_{i}U_{ss}n_{si\uparrow}n_{si\downarrow}+U_{sp}\left(n_{pi\uparrow}+n_{pi\downarrow}\right)\left(n_{si\uparrow}+n_{si\downarrow}\right)}_{\approx H_{\rm el-el}^{(0)}},\label{eq:Hprime1}
\end{equation}
where $\nu\in \{s,p\}$. In principle, the tunneling $t_p$ of the p-orbital can be anisotropic along different directions, but for simplicity we will take an isotropic value here, with the value determined by Sec.~\ref{subsec:scales}. In the large $d/a_0$ limit, we approximate $\epsilon_{s,p}$ by the corresponding energy levels of the hydrogen atom. This also assumes that there is one electron per site, as detailed in Sec.~\ref{subsec:electrostatic}.

In our limit of interest of large $d/a_0$ and half filling, the on-site interaction terms $U_{ss},U_{sp}$ originating from $H_{\rm el-el}^{(0)}$ greatly exceed the tunneling rates. The Hilbert space thus separates into a low-energy manifold consisting of one electron per site, and a high-energy manifold where two electrons occupy the same nucleus. Using a Schrieffer-Wolff transformation~\cite{bravyi_schrieffer-wolff_2011}, one can project the dynamics under $H'$ into the low-energy manifold. In particular, we define $P$ as the projector into the manifold of one electron per nucleus and $Q=1-P$ as its complement. The Schrieffer-Wolff transformation then states that the effective low-energy Hamiltonian is given by $H'_{\rm eff}=\frac{1}{2}[S,H_{\rm el}]$, where
\begin{equation} S=\sum_{n,n'}\frac{\langle n|PH_{\rm el}Q+QH_{\rm el}P|n'\rangle}{E_n-E_{n'}}|n\rangle\langle n'|.\label{eq:SW} \end{equation}
Here, $|n\rangle,|n'\rangle$ are eigenstates of $H_{\rm el-el}^{(0)}$ and $E_{n,n'}$ the corresponding energies, while $H_{\rm el}=H_{\rm el}^{(0)}+H_{\rm el}^{(1)}$ is the single-electron Hamiltonian limited to on-site and nearest neighbor terms. Evaluation of this equation gives
\begin{equation}
    H'_{\rm eff}=H_{\rm tJ}-\frac{4t_p^2}{U_{sp}}\sum_{i\sigma}n_{pi\sigma}-\frac{t_s^2}{U_{sp}}\sum_{\langle ij\rangle \sigma\sigma'}\left(n_{si\sigma}n_{pj\sigma'}+n_{sj\sigma}n_{pi\sigma'}\right).
\end{equation}
The term proportional to $n_{pi\sigma}$ describes a~(small) overall renormalization of the p-orbital energy, which is not relevant to our discussion. Furthermore, in the limit of weak driving by an external field such that only one p-orbital is excited~(linear optical response), the last term proportional to $t_s^2/U_{sp}$ has no effect on the dynamics. Indeed, this term describes a conditional shift of the neighboring s-orbital energies~(and thus their transition energies), once a p-orbital is already excited, and thus only contributes a nonlinear optical effect. We have thus derived the $H_{\rm tJ}$ Hamiltonian that appears in Eq.~(\ref{eq:H2D}). 

It can be noted that $H'$ in Eq.~(\ref{eq:Hprime1}) is essentially a two-band Fermi-Hubbard model. Certainly, the Fermi-Hubbard model over-simplifies the full quantum chemistry problem of an array of hydrogen atoms. Perhaps most prominently, the on-site interaction energies $U$, as estimated from the hydrogen electron wave functions, are on the order of the ionization energy of hydrogen itself, which implies that higher bands are needed to accurately reproduce the full electronic wave functions of the array. Nonetheless, state-of-the-art computational quantum chemistry calculations~\cite{simons_collaboration_on_the_many-electron_problem_ground-state_2020} on the ground state of a 1D hydrogen chain at large lattice constants suggest that the Fermi-Hubbard model does describe well the key physics~(\textit{e.g.}, spin correlations consistent with the Heisenberg spin model). Although such a direct comparison in 2D is beyond numerical capabilities, we take the 1D results as sufficient justification for the reduction to the 2D Fermi-Hubbard model.

\subsection{Density-density correlations: holon-doublon pairs}
The intermediate states in perturbation theory that give rise to the Heisenberg spin interaction $H_J$~(see Fig.~\ref{fig:2Dexplained}a and encoded in Eq.~(\ref{eq:SW})) describe the onset of density-density correlations in the ground state, due to Coulomb interactions between electrons. Specifically, these states describe holon-doublon pairs, bound states consisting of an empty nucleus and a doubly occupied one~(Fig.~\ref{fig:2Dexplained}b). They will influence the optical response of a 2D array, and thus we summarize how their population in the many-body ground state can be calculated via a slave-fermion formalism introduced in Ref.~\cite{han_charge_2016}. In this approach, one amplifies the Hilbert space associated with s-orbitals on each site, defining bosonic operators $s_{i\sigma}^{\dagger}$ that create a spin-$\sigma$ ``spinon'' particle on each site, and fermionic operators $e_i^{\dagger}$ and $d_i^{\dagger}$ that create ``holon'' and ``doublon'' particles, respectively. The physical fermion operator can be expressed in terms of these new particles as
\begin{align}
\label{eq:slave-fermion1}
    b_{si\sigma}=s_{i\overline{\sigma}}^\dagger d_i+\mathrm{sign}(\sigma)s_{i\sigma}e_i^\dagger\,,
\end{align}
where $\overline{\sigma}$ denotes the opposite spin value of $\sigma$, and $\mathrm{sign}(\uparrow)=-\mathrm{sign}(\downarrow)=1$. The physical Hilbert space is preserved by imposing the population constraint on the new particles,
\begin{align}
\label{eq:slave-fermion2}
   e_i^\dagger e_i+d_i^\dagger d_i+\sum_\sigma s_{i\sigma}^\dagger s_{i\sigma}=1\,.
\end{align}
While the discussion thus far just amounts to a formal re-mapping of the problem, a considerable simplification arises by assuming spin-charge separation. In particular, in the limit $|t_s|/U_{ss}\ll 1$, the spin sector $s_{i\sigma}$ is assumed to be characterized by the ground state of the Heisenberg interaction term $H_J=J\sum_{\langle ij\rangle} \bm{S}_i\cdot \bm{S}_j$ of the tJ model within linear spin wave theory~\cite{auerbach_interacting_1994}, which is known to reproduce accurately key quantities like the sub-lattice magnetization~\cite{manousakis_spin-12_1991}. The properties of the charge sector involving holons and doublons can then be computed taking into account the coupling of the chargons to the background spin waves, while assuming that the charge dynamics do not impart back-action on the spin sector. This approximate theory is known to exhibit good agreement with state-of-the-art numerical computations on the half-filled Fermi-Hubbard model~\cite{han_charge_2016}.

Specifically, within linear spin wave theory, it is first assumed that at the mean-field level, there exists anti-ferromagnetic N\'{e}el order. This is captured by dividing the square lattice into alternating sub-lattices A and B, and assigning a mean-field value $s_{i\uparrow},s_{i\uparrow}^\dagger\to s_0$ and $s_{i\downarrow},s_{i\downarrow}^\dagger\to s_0$ to the spins on sub-lattices A and B, respectively. The amplitude $s_0\in\mathbb{R}$ can eventually be computed by taking the expectation value of Eq.~\eqref{eq:slave-fermion2}, whereby
\begin{align}
\label{eq:slave-fermion3}
    s_0^2=\begin{cases}
        1-\langle e_i^\dagger e_i\rangle-\langle d_i^\dagger d_i\rangle-\langle s_{i\downarrow}^\dagger s_{i\downarrow}\rangle & i\in A \\
        1-\langle e_i^\dagger e_i\rangle-\langle d_i^\dagger d_i\rangle-\langle s_{i\uparrow}^\dagger s_{i\uparrow}\rangle & i\in B
    \end{cases}
\end{align}
and where $\langle\dots\rangle$ indicates the expectation value in the many-body ground state. This allows to implement the number conservation constraint self-consistently~\cite{han_charge_2016}. 

One can then re-write the Heisenberg Hamiltonian $H_J$ in terms of these mean-field values and the remaining spinon operators  $s_{i\downarrow},s_{i\uparrow}$ describing fluctuations on sub-lattices A and B, respectively. This Hamlitonian can be diagonalized in momentum space, as
\begin{align}
\label{eq:magnonHamiltonian}
    H_J=2Js_0^2\sum_\bfk \Omega_\bfk c_\bfk^\dagger c_\bfk\,,
\end{align}
where $\Omega_\bfk=\sqrt{1-\gamma_\bfk^2}$ with $\gamma_\bfk=(\cos k_xd+\cos k_yd)/2$ and with bosonic Bogoliubov operators $c_\bfk=u_\bfk s_\bfk+v_\bfk s_{-\bfk}^\dagger$ of the Fourier transformed spinon operators $s_\bfk$ on either sub-lattice, with $u_\bfk^2-v_\bfk^2=1$ and $u_\bfk v_\bfk=-\gamma_\bfk/2\Omega_\bfk$. The spin sector thus supports a single band of low-lying collective magnon excitations, created by $c_\bfk^\dagger$.

We now return to the single-band Fermi-Hubbard model, as described by the terms in Eq.~\eqref{eq:Hprime1} involving only s-orbital electrons. Re-writing this Hamiltonian in terms of the spinon and chargon operators gives 
\begin{align}
\label{eq:fullHubbard}
\begin{split}
    H'
    &=2Js_0^2\sum_\bfk \Omega_\bfk c_\bfk^\dagger c_\bfk+\frac{U_{ss}}{2}\sum_\bfk\left(d_\bfk^\dagger d_\bfk+e_\bfk^\dagger e_\bfk\right) \\
    &\qquad+4t_ss_0^2\sum_\bfk\left(d_{\mathbf{Q}-\bfk}^\dagger e_\bfk^\dagger+h.c.\right)+4t_ss_0\sum_{\bfk,\bfq} V_{\bfk\bfq}\left(d_\bfk^\dagger d_{\bfk+\bfq}c_\bfq^\dagger+e_{\bfk+\bfq}e_\bfk^\dagger c_\bfq^\dagger+h.c.\right)\,,
\end{split}
\end{align}
where we have defined $\mathbf{Q}=(\pi/d,\pi/d)^T$ and $V_{\bfk\bfq}=-(\gamma_{\bfk+\bfq}u_\bfq+\gamma_\bfk v_\bfq)/\sqrt{N}$. (Here, and also in following sub-sections, we will simply re-use the notation $H'$ to avoid defining excessive new variables, with the understanding that the definitions of $H'$ in different sub-sections are distinct.) Physically, the first two terms in this Hamiltonian are the spinon and chargon self-energies and the third term describes the creation and annihilation of holon-doublon pairs from the mean-field spin background. The fourth term represents a coupling of the chargons to the background spin fluctuations.

To calculate the holon and doublon populations at the onset of quantum chemistry (i.e. at large $d/a_0$ or equivalently at lowest order in $t_s/U_{ss}$), the last term in Eq.~\eqref{eq:fullHubbard} can be dropped and one gets a free particle theory in the charge degrees of freedom. The free particle Hamiltonian can be diagonalized by a fermionic Bogoliubov transformation, and evaluating Eq.~\eqref{eq:slave-fermion3} then yields $s_0^2\approx0.803$. 
In a similar manner, we calculate the number of holon-doublon pairs $P_{\rm hd}$ over the total number of lattice sites, as $P_{\rm hd}= \sum_i\langle d_i^\dagger d_i \rangle/N$, which straightforwardly gives
\begin{align}
\label{eq:hd_population}
    P_{\rm hd}\approx \frac{1}{N}\left(\frac{t_s}{U_{ss}}\right)^2\sum_{\bfk\in\mathrm{1BZ}}(4s_0^2\gamma_\bfk)^2,
\end{align}
with the sum evaluated over the first Brillouin zone as indicated. Then, in the thermodynamic limit, 
$P_{\rm hd}\approx 2.58(t_s/U_{ss})^2$ as stated in Sec.~\ref{sec:model}.

\subsection{Electrostatic interactions}\label{subsec:electrostatic}

Here, we analyze in more detail the Coulomb interactions both between electrons and nuclei, and between electrons, with the goal of deriving $H_0$ and the so-called longitudinal part of $H_{\rm dip-dip}$ in Eq.~(\ref{eq:H2D}). The on-site term of the single-particle Hamiltonian $H_{\rm el}^{(0)}$ is band-diagonal and can be written as $H_{\rm el}^{(0)}=\sum_{i\sigma\nu}\epsilon_\nu n_{\nu i \sigma}$, where
\begin{equation}
    \epsilon_{\nu} = \int d^3\bfr~  \phi^{\ast}_{\nu}(\bfr)\left(\frac{\bfp^2}{2m}-\sum_j V_C(\bfr+\bfR_{ij})\right)\phi_{\nu}(\bfr).
\end{equation}
In the large $d/a_0$ limit, it is convenient to expand the Coulomb potential $V_C(\bfr-\bfR_{ij})$ as a multipolar expansion. Explicitly,
\begin{equation}
\label{eq:Hel-expansion}
    \epsilon_{\nu} = \int d^3\bfr~  \phi^{\ast}_{\nu}(\bfr)\left(\frac{\bfp^2}{2m}- V_C(\bfr)\right)\phi_{\nu}(\bfr)-\frac{q^2}{4\pi\varepsilon_0}\sum_{j\neq i}\frac{1}{R_{ij}}-\frac{q}{4\pi\varepsilon_0}\sum_{j\neq i}\frac{\bfR_{ij}\cdot\bfq_{\nu\nu}\cdot\bfR_{ij}}{R_{ij}^5}+\dots\,,
\end{equation}
where we have defined $\bfR_{ij}=\bfR_i-\bfR_j$ and $R_{ij}=|\bfR_{ij}|$ as well as the quadrupole moment $\bfq_{\nu\nu'}=q\int d^3\bfr~\phi_\nu^*(\bfr)(3\bfr\otimes\bfr-|\bfr|^2)\phi_{\nu'}(\bfr)$. We have also used the parity of the Wannier functions to conclude that $\bfq_{\nu\nu'}$ is diagonal in the band indices~(with $\nu,\nu'$ restricted to $s,p$) and to discard a first-order dipolar term. It should be noted that $\epsilon_{\nu}$ diverges due to the sum $\sum_{j\neq i}R_{ij}^{-1}$, which reflects the infinite potential energy of a point-like electron within a lattice of positive charges. At half filling, however, each nucleus is overall charge neutral due to the presence of an electron at the same site, eliminating this infinity. Formally, the divergence cancels with a corresponding term in $H_{\rm el-el}^{(1)}$.

To establish this, we now consider the electron-electron interactions $H_{\rm el-el}^{(1)}$ involving pairs of sites~($i=l$ and $j=k$ in Eq.~(\ref{eq:Wannier_expansion})). Applying a similar multipolar expansion,
\begin{equation}
\begin{split}
     H^{(1)}_{\rm el-el}
     &=\frac{q^2}{8\pi\varepsilon_0}\sum_{i,j\neq i,\nu\nu',\sigma\sigma'}\frac{1}{R_{ij}}n_{\nu i\sigma}n_{\nu' j\sigma'}+\frac{q}{4\pi\varepsilon_0}\sum_{i,j\neq i,\nu\nu',\sigma\sigma'}\frac{\bfR_{ij}\cdot\bfq_{\nu\nu}\cdot\bfR_{ij}}{R_{ij}^5}n_{\nu i\sigma}n_{\nu' j\sigma'} \\
     &\qquad\qquad+\sum_{i,j\neq i,\nu\mu,\sigma\sigma'}\frac{\vecTemp{d}_{\mu\overline{\mu}}\cdot(1-3\hat{\bfR}_{ij}\otimes\hat{\bfR}_{ij})\cdot\vecTemp{d}_{\nu\overline{\nu}}}{8\pi\varepsilon_0R_{ij}^3}\left(b_{\mu i\sigma}^\dagger b_{\overline{\mu}i\sigma}\right)\left(b_{\nu j\sigma'}^\dagger b_{\overline{\nu}j\sigma'}\right)+\dots\,,
\end{split}
\end{equation}
where we have defined $\hat{\bfR}_{ij}=\bfR_{ij}/R_{ij}$ and the dipole moment $\vecTemp{d}_{\nu\nu'}=q\int d^3\bfr~\bfr~\phi_\nu^*(\bfr)\phi_{\nu'}(\bfr)$, and where $\overline{\nu}=p$~$(s)$ when $\nu=s$~$(p)$. The first two terms in this expression will cancel with the terms in the multipolar expansion of $H_{\rm el}^{(0)}$ under the assumption of one electron per site, $\sum_{\nu\sigma}n_{\nu i\sigma}\approx 1$. This justifies keeping only the first term on the right hand side of Eq.~(\ref{eq:Hel-expansion}) for $\epsilon_{\nu}$. We furthermore approximate this first term by the energy eigenvalues of the bare hydrogen atom.

The last term in the expansion of $H^{(1)}_{\rm el-el}$ is not cancelled, however. Physically, it drives  transitions between the $s$- and $p$-like orbitals on pairs of sites. The net electrostatic interaction is therefore captured, to first-order in the expansion in $d/a_0$, by the Hamiltonian
\begin{equation}
     H'
     =\sum_{ij\neq i,\nu\mu,\sigma\sigma'}\frac{\vecTemp{d}_{\mu\overline{\mu}}\cdot(1-3\hat{\bfR}_{ij}\otimes\hat{\bfR}_{ij})\cdot\vecTemp{d}_{\nu\overline{\nu}}}{8\pi\varepsilon_0R_{ij}^3}\left(b_{\mu i\sigma}^\dagger b_{\overline{\mu}i\sigma}\right)\left(b_{\nu j\sigma'}^\dagger b_{\overline{\nu}j\sigma'}\right)\,.
\end{equation}
Some terms in $H'$ are energy non-conserving (\textit{e.g.}, two s-orbitals being annihilated to form two p-orbitals). These encode the well-known van der Waals interaction, where fluctuations involving these interactions lead within perturbation theory to a $\sim 1/R_{ij}^6$ attractive potential between two ground-state atoms. Of relevance to us at large $d/a_0$ are the resonant terms in $H'$, which show the characteristic $\sim 1/R_{ij}^3$ scaling of near-field dipole-dipole interactions and enable energy transfer. In the notation of Sec.~\ref{subsec:Formalism}, $\vecTemp{d}_{sp}=p_0\hat{\vecTemp{x}}$ and therefore $H'$ takes the effective form
\begin{equation}
    H_{\rm eff}'
    =-\Gamma_0\sum_{ij\neq i,\sigma\sigma'}G_{ij}^\parallel\left(b_{s i\sigma}^\dagger b_{pi\sigma}\right)\left(b_{p j\sigma'}^\dagger b_{sj\sigma'}\right)\,.\label{eq:Glong}
\end{equation}
in terms of the single-atom spontaneous decay rate $\Gamma_0$ introduced previously. We see that this encodes the part of the dipole-dipole interaction Hamiltonian $H_{\rm dip-dip}$ in Eq.~(\ref{eq:3DQO_Hamiltonian}) associated with the longitudinal component of the Green's function. In particular, $G_{ij}^\parallel=\hat{\vecTemp{x}}\cdot\vecTemp{G}^\parallel(\bfR_{ij},\omega_0)\cdot\hat{\vecTemp{x}}$ with $\nabla\times\vecTemp{G}^\parallel(\bfr,\omega_0)=0$. Explicitly,
\begin{equation}
\label{eq:G_L_full_equation}
 \vhat{x} \cdot\vecTemp{G}^\parallel(\bfr,\omega_0)\cdot \vhat{x} =
\dfrac{3}{4}
\left[-\dfrac{1}{(k_0 |\bfr|)^3}+\dfrac{3}{(k_0 |\bfr|)^3}\dfrac{ (\hat x\cdot \bfr)^2  }{|\bfr|^2}\right]\,.
\end{equation}

\subsection{Photon-mediated interactions}

The remaining transverse component of the dipole-dipole interactions in $H_{\rm dip-dip}$ originates from integrating out the photons in $H_{\rm ph-el}$ in Eq.~(\ref{eq:H_longwavelength}), and restricted to the terms describing on-site electronic transitions driven by the field. In particular, we consider the minimal light-matter Hamiltonian $H'=H_0+H_{\rm ph}+H_{\rm ph-el}^{\rm (0)}$, where 
\begin{equation}
    H_{\rm ph-el}^{(0)} = -i\omega_0p_0\sum_{i\sigma}\vecTemp{A}(\bfR_i)\cdot\hat{\vecTemp{x}}\left(b_{p i\sigma}^{\dagger}b_{s i\sigma}-h.c.\right)\,.  \label{eq:H_int}
\end{equation}
Above, we have used the parity of the Wannier functions and the relation $\langle\phi_s(\bfr-\bfR_i)|\bfp|\phi_p(\bfr-\bfR_i)\rangle=im\omega_0\bfp_0/q$ for the momentum matrix element, where $\vecTemp p_0$ is the dipole matrix element defined in Sec.~\ref{subsec:Formalism}. Physically, this Hamiltonian states that transitions between $s$ and $p$ are accompanied by photon emission/absorption. Emitted photons can then subsequently propagate and either induce transitions in other atoms, or propagate beyond the atomic medium altogether, resulting in spontaneous emission and energy loss of the atomic subsystem.

Formally, one can integrate out the photons and derive the dynamics of the reduced atomic density matrix $\rho(t)$ within the standard Born-Markov approximation, to obtain
\begin{equation}
    \dot{\rho}(t)\approx-\frac{i}{\hbar}[H_0,\rho(t)]-\frac{1}{\hbar^2}\int_0^\infty dt'\langle e^{-itH_0/\hbar}[V_{\rm ph-el}(t),[V_{\rm ph-el}(t-t'),\rho(t)\otimes |0\rangle\langle 0|]]e^{itH_0/\hbar}\rangle \,, \label{eq:master_eqn}
\end{equation}
where $V_{\rm ph-el}(t)=e^{it(H_0+H_{\rm ph})/\hbar}H_{\rm ph-el}^{(0)}e^{-it(H_0+H_{\rm ph})/\hbar}$ represents the interaction Hamiltonian transformed to the interaction picture and $\langle\dots\rangle$ denotes an expectation value with respect to the field vacuum $|0\rangle$. Evaluating Eq.~(\ref{eq:master_eqn}) gives rise to a master equation of the form $\dot{\rho}(t)=-i/\hbar(H_{\rm eff}'\rho(t)-\rho(t)H_{\rm eff}'^{\dagger})+\sum_i J_{i}\rho(t)J_{i}^{\dagger}$. Within the quantum jump formalism, this equation describes time evolution under a non-Hermitian Hamiltonian $H_{\rm eff}'$ punctuated stochastically by quantum jumps realized by a set of jump operators $J_i$, whose explicit form is not relevant here. Explicitly, 
\begin{equation}
    H'_{\rm eff}=H_0-\frac{i}{\hbar}\int_0^\infty dt'\langle e^{-itH_0/\hbar}V_{\rm ph-el}(t)V_{\rm ph-el}(t-t')e^{itH_0/\hbar}\rangle \,.
\end{equation}
Substituting the explicit expression for $V_{\rm ph-el}(t)$ implied by Eq.~\eqref{eq:H_int}, the effective Hamiltonian can be written in terms of the time-correlator $\bm{C}_{ij}(t)=\langle \vecTemp{A}(\bfR_i,t)\otimes \vecTemp{A}(\bfR_j,0)\rangle$ of the interaction picture vector potential operators as
\begin{equation}
    H'_{\rm eff}=H_0-\frac{i\omega_0^2p_0^2}{\hbar}\sum_{ij,\sigma\sigma'}\int_0^\infty dt'\left[\hat{\vecTemp{x}}\cdot\bm{C}_{ij}(t')\cdot\hat{\vecTemp{x}}\left(b_{pi\sigma}^{\dagger}b_{si\sigma}\right)\left(b_{sj\sigma'}^{\dagger}b_{pj\sigma'}\right)e^{i\omega_0t'}+h.c.\right]\,.
    \label{eq:effective H_ph-el 1}
\end{equation}
Above, we have only retained energy-conserving terms, where the photon mediates a de-excitation and excitation of a p- and s-orbital, respectively. The off-resonant terms, on the other hand, are the photon-mediated counterparts to the van der Waals potential of the electrostatic interaction, which are often referred to as the Casimir-Polder potential~\cite{buhmann_dispersion_2007}. These  produce an overall shift of ground state energy between a collection of atoms in their s-orbitals, which is not relevant for our purposes.

The correlator $\bm{C}_{ij}(t)$ of the vacuum field can readily be calculated to yield
\begin{equation}
    H'_{\rm eff}=H_0-\Gamma_0\sum_{ij\neq i,\sigma\sigma'}G_{ij}^\perp\left(b_{s i\sigma}^\dagger b_{pi\sigma}\right)\left(b_{p j\sigma'}^\dagger b_{sj\sigma'}\right)\,,
    \label{eq:effective H_ph-el 2}
\end{equation}
where now $G^{\perp}_{ij}=G_{ij}-G_{ij}^{\parallel}$ is the projection along $x$ of the transverse part of the Green's function~($\nabla\cdot\vecTemp{G}^\perp(\bfr,\omega_0)=0$). Combining this result with Eq.~(\ref{eq:Glong}) gives the total dipole-dipole interaction $H_{\rm dip-dip}$ of Eq.~(\ref{eq:3DQO_Hamiltonian}).

\section{Refractive index: the quantum chemistry limit}\label{sec:indexchemistry}

In Sec.~\ref{sec:model}, we presented our effective model and Hamiltonian~(\ref{eq:H2D}) to describe atom-light interactions, including non-perturbative multiple scattering, and quantum chemistry within a 2D array at large lattice constant $d/a_0$. Three main effects that emerge from chemistry are quantum magnetism, electronic density-density correlations, and hopping dynamics of photo-excited electrons. Here, we analyze their effects in limiting the refractive index of a 3D crystal.

\begin{figure}[t]
\centering
\includegraphics[width=0.85\textwidth]{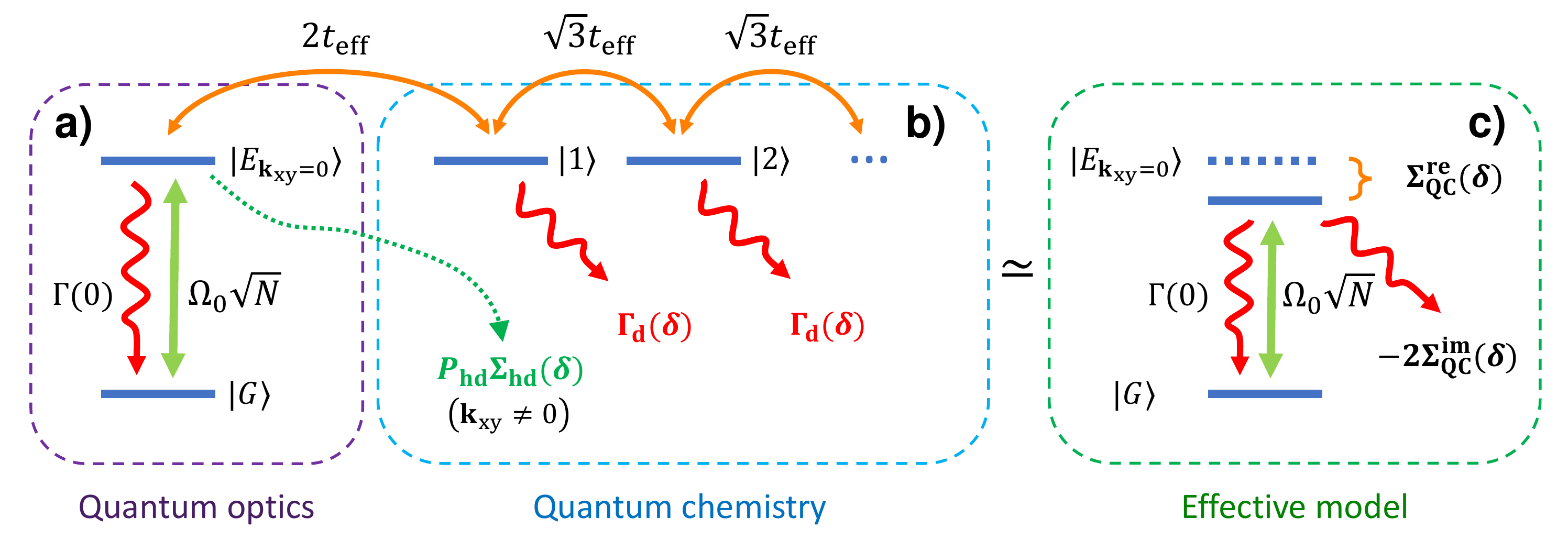}
\caption{a) In the quantum optics regime, a large refractive index is achieved due to the single-mode response of an individual 2D layer, where weak incident light only couples the many-electron ground state $|G\rangle$ to a single collective excited state $|E_{\bfk_{xy}=0}\rangle$, and this state emits elastically back into the same optical mode at a rate $\Gamma(0)$. b) Quantum chemistry allows for inelastic or spatial multi-mode emission. Spatial multi-mode emission into directions $\bfk_{xy}\neq 0$ arises from light scattering off of electronic density-density correlations, in the form of holon-doublon pairs~(dashed green arrow). Hopping of the photo-excited electron at a rate $\sim t_{\rm eff}$ couples the collective excited state $|E_{{\bf k}_{xy}=0}\rangle$ to a continuum of additional states $|n\rangle$, labeled by an integer $n$ that describes the degree to which the hopping disturbs the anti-ferromagnetic N\'{e}el order of the electron spins, as described further in the main text. This process, along with the effective decay rate $ \Gamma_{\rm d}(\delta)=-2\im\,\chi(0,\delta)^{-1}$ of the excited electron, leads to inelastic emission. c) The various quantum chemistry processes illustrated in Fig.~\ref{fig:chemistry_optics}b give rise to a modified optical response of the 2D layer, which can be captured by a complex self-energy $\Sigma_{\rm QC}(\delta)$ of the excited state $|E_{\bfk_{xy}=0}\rangle$. The real and imaginary parts describe an chemistry-induced energy shift and effective inelastic decay rate, respectively.
}
\label{fig:chemistry_optics}
\end{figure}

We begin by recalling the main result in the quantum optics limit of Sec.~\ref{sec:QO}, involving only the $H_{\rm QO}$ term of Eq.~(\ref{eq:H2D}). In particular, for weak light at normal incidence, a 2D array behaves as a single-mode system, where the light excites only a single collective mode $|E_{\bfk_{xy}=0}\rangle$, \textit{and} this collective mode only re-radiates light elastically at a rate $\Gamma(\bfk_{xy}=0)$ back in the same $\bfk_{xy}=0$ direction~(either forward or backward). This single-mode nature of the quantum optics limit is illustrated in Fig.~\ref{fig:chemistry_optics}a~(dashed purple box), and produces the large, purely real refractive index of a 3D lattice. One effect that emerges due to quantum chemistry is the appearance of anti-ferromagnetic N\'{e}el ordering in the spin component $|\sigma\rangle$ of the many-electron ground state $|G\rangle$, as discussed in Sec.~\ref{subsubsec:presentation}. This same spin wave function is inherited by $|E_{\bfk_{xy}=0}\rangle$, as the exciting light does not affect spin. The N\'{e}el ordering in itself thus does not alter the refractive index.

In contrast, the excited electron dynamics and density-density correlations break the single mode response by allowing for inelastic or spatial multi-mode emission processes, as illustrated in Fig.~\ref{fig:chemistry_optics}b~(dashed blue box). We now describe their optical effects in greater detail, and describe how they can be incorporated into a frequency-dependent effective level shift and inelastic decay rate of the excited state, as characterized respectively by the real and imaginary parts of a self-energy term $\Sigma_{\rm QC}(\delta)$~(Fig.~\ref{fig:chemistry_optics}c).

\subsection{Dynamics of photo-excited electron}
\label{subsec:photo_excited_dynamics}

\begin{figure}[t]
\centering
\includegraphics[width=0.8\textwidth]{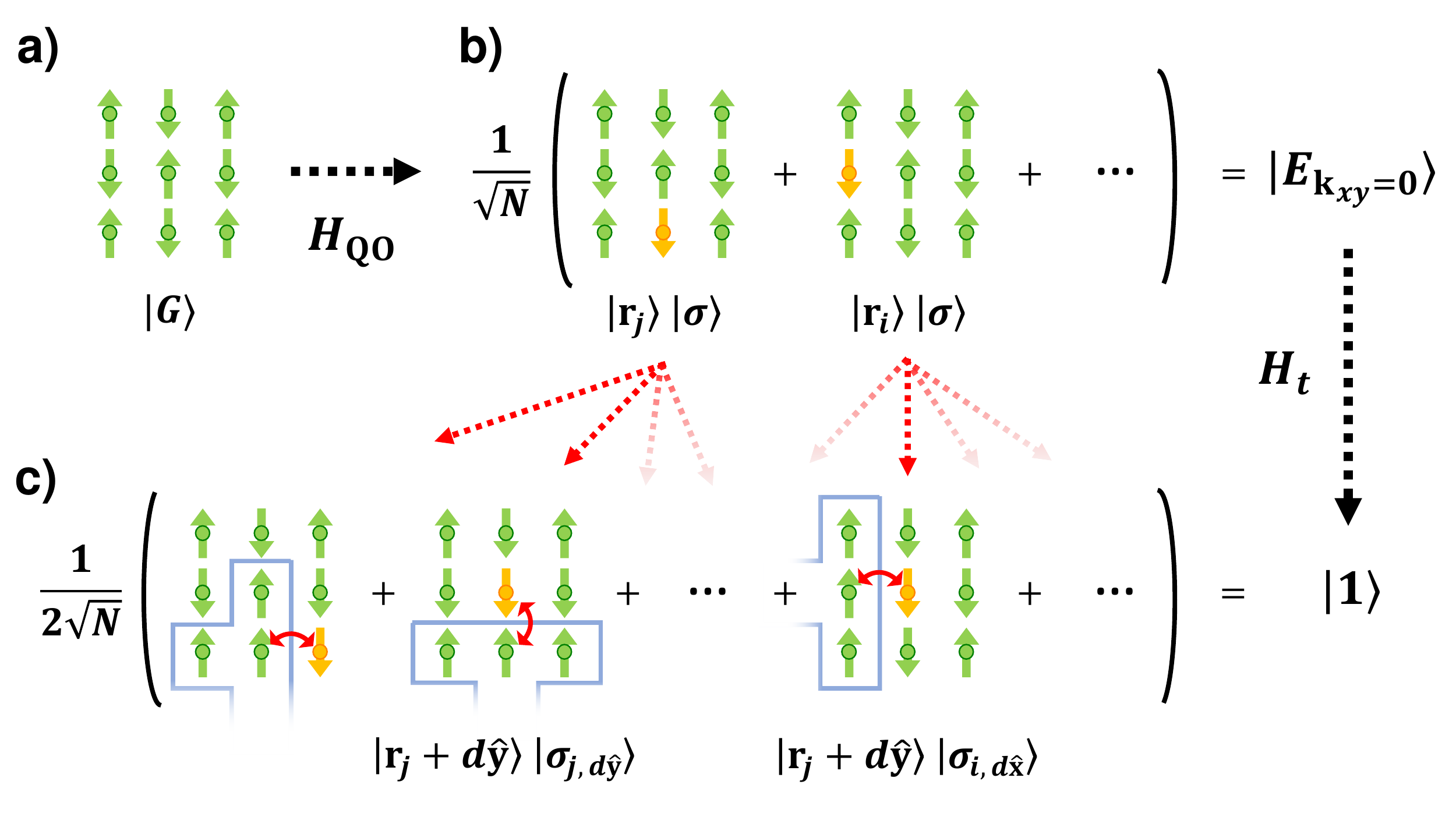} 
\caption{Physics of the ${\rm tJ}_z$ model. a) The many-body ground state of the 2D array consists of a single s-orbital electron per site~(green), with classical anti-ferromagnetic N\'{e}el order of the spins, as indicated by the arrows. b) A single incident photon excites the array into the superposition state $|E_{\bfk_{xy}=0}\rangle$, where any electron at site $j$ is equally excited to a p-orbital~(yellow) without changing the spin. c) Dynamics of the excited electron via the Hamiltonian $H_t$ couples the state $|E_{\bfk_{xy}=0}\rangle$ to the state $|1\rangle$, a superposition state of all possibilities where the p-orbital can exchange orbital and spin degrees of freedom with an s-orbital electron in a neighboring site~(solid red arrows). This dynamics breaks the perfect N\'{e}el order for the s-orbital electrons inside the blue boxes. Certain configurations making up the superposition are labeled and described further in the main text.
}
\label{fig:tJ}
\end{figure}
In this subsection, we neglect density-density correlations~(i.e. assuming exactly one electron per site), and focus on the effect of photo-excited electron dynamics as described by the tJ model Hamiltonian $H_{\rm tJ}$ in Eq.~(\ref{eq:H2D}). Instead of dealing directly with $H_{\rm tJ}$, we will work with the simpler tJ${}_z$ model, 
which 
is known to capture well the dynamics at short times~\cite{golez_mechanism_2014}. In the tJ${}_z$ model, only the $z$ components of the spins are assumed to interact, 
\begin{equation}
    H_{{\rm tJ}_z}=\underbrace{J\sum_{\langle ij\rangle} S_{iz}S_{jz}}_{H_{J_z}} \underbrace{- t_{\rm eff}\sum_{\langle ij\rangle \sigma\sigma'} (b_{pi\sigma}^{\dagger}b_{pj\sigma}b_{sj\sigma'}^{\dagger}b_{si\sigma'}+h.c.)}_{H_t},
\end{equation}
where again the spin interaction $J$ is restricted to electrons in s-orbitals. In this case, classical antiferromagnetic N\'{e}el order describes exactly the global spin ground state configuration $|\sigma\rangle$ of the electronic ground state $|G\rangle$ and excited state $|E_{\bfk_{xy}=0}\rangle$, as illustrated in Figs.~\ref{fig:tJ}a,b. The excited state is an equal superposition of the excited p-orbital being located at different sites, as we qualitatively show in Fig.~\ref{fig:tJ}b by depicting two representative configurations in the overall superposition.

We thus want to derive the index of the system evolving under Eq.~(\ref{eq:H2D}) and with the replacement $H_{\rm tJ}\rightarrow H_{\rm tJ_z}$. We also assume that due to the small magnitude of $J$~(compared to both $\Gamma(0)$ and $t_{\rm eff}$, as seen in Fig.~\ref{fig:parameter_estimates}b), the energies $E_{\sigma}$ of different spin configurations will be non-zero but negligible from the standpoint of phase evolution $e^{-iE_{\sigma}t}$. Furthermore, we will ignore the contributions of $H_{\rm drive}$ beyond the matrix element connecting $|G\rangle$ and $|E_{\bfk_{xy}=0}\rangle$, as all other contributions only lead to multi-photon corrections in the refractive index that are nonlinear in the field intensity.

The first non-trivial effect beyond the quantum optics limit arises from $H_t$ acting on $|E_{\bfk_{xy}=0}\rangle$. As illustrated in Fig.~\ref{fig:tJ}b, the state $|E_{\bfk_{xy}=0}\rangle=N^{-1/2}(\sum_j |\bfr_j\rangle)|\sigma\rangle$ can be expressed as an equal-weight superposition, where $|\bfr_j\rangle$ denotes that the excited p-orbital is located at site $\bfr_j$ and $|\sigma\rangle$ is the ground-state spin configuration. $H_t$ allows the excited electron to exchange both its orbital and spin degrees of freedom with any nearest neighbor, thus $H_t$ couples $|E_{\bfk_{xy}=0}\rangle$ to the new normalized state
\begin{equation} 
|1\rangle=\frac{1}{2\sqrt{N}}\sum_j\sum_{{\bm \delta}_1=\pm d\vhat{x},\pm d\vhat{y}} |\bfr_j+{\bm \delta}_1\rangle|\sigma_{j,{\bm \delta}_1}\rangle,\label{eq:1} \end{equation}
where $|\bfr_j+{\bm \delta}_1\rangle$ describes the position of the p-orbital following a move in the nearest neighbor direction ${\bm \delta}_1$ and $|\sigma_{j,{\bm \delta}_1}\rangle$ is the spin state following the corresponding spin exchange. One sees that the states $|\sigma_{j,{\bm \delta_1}}\rangle$ break the perfect N\'{e}el order, as indicated by the blue boxes. From these boxes, one also visualizes that all spin states are orthogonal to one another, $\langle \sigma_{j,{\bm \delta}_1}|\sigma_{j',{\bm \delta}'_1}\rangle=\delta_{j,j'}\delta_{{\bm \sigma}_1,{\bm \sigma}'_1}$ and thus the state $|1\rangle$ is entangled in the orbital and spin degrees of freedom. The matrix element of the interaction is $\langle 1|H_t|E_{\bfk_{xy}=0}\rangle=-2t_{\rm eff}$. 

A key consequence of the above discussion is that the dynamics of $H_t$ results in distinguishable spin backgrounds, even when the p-orbital winds up on the same final site. This is illustrated in Fig.~\ref{fig:tJ}b and c, where on one hand in Fig.~\ref{fig:tJ}b we explicitly show two positions $|\bfr_j\rangle$ and $|\bfr_i=\bfr_j-d\vhat{x}+d\vhat{y}\rangle$ of the p-orbital in the state $|E_{\bfk_{xy}=0}\rangle$, and on the other hand in Fig.~\ref{fig:tJ}c we draw the new orbital states $|\bfr_j+d\vhat{y}\rangle$ following an upward move and $|(\bfr_j-d\vhat{x}+d\vhat{y})+d\vhat{x})\rangle=|\bfr_j+d\vhat{y}\rangle$ following a rightward move, respectively. 
Despite the orbital wave functions being the same, the orthogonality of the associated spin wave functions $\ket{\sigma_{j,d\vhat y}}$ and $\ket{\sigma_{i,d\vhat x}}$ is seen by the different blue boxes indicating where the N\'{e}el order has been broken as a result of the p-orbital motion. Note that should broken order be left behind once the p-orbital relaxes by photon emission, the photon emission will be inelastic and thus contributes an imaginary component to the refractive index.

To calculate the effect on the index, we must understand how state $|1\rangle$ further evolves under excited-state hopping dynamics and dipole-dipole interactions, as contained in the approximate Hamiltonian $H_t+H_0+H_{\rm dip-dip}$~(from above, recall that we ignore $H_{J_z}$ and $H_{\rm drive}$ in subsequent evolution). These processes are pictorially described in \figref{fig:chemistry_optics}b, by orange ($H_t$) arrows denoting further hopping dynamics, and red, wavy arrows ($H_0+H_{\rm dip-dip}$) denoting dipole-dipole interactions. Due to the different scalings of the interactions seen in Fig.~\ref{fig:parameter_estimates}b, we consider simpler limits where either $H_t$ or $H_0+H_{\rm dip-dip}$ completely dominates. In any case, from the standpoint of $|E_{\bfk_{xy}=0}\rangle$, these dynamics couple this state to a continuum. This leads to an effective decay rate other than the preferred elastic emission channel, decreasing the optical response. Our goal is to quantify this in terms of a ``self-energy'' contribution to state $|E_{\bfk_{xy}=0}\rangle$.

We first consider when $H_0+H_{\rm dip-dip}$ dominates subsequent evolution of $|1\rangle$. Since $H_0+H_{\rm dip-dip}$ does not couple to spins, the various states in $|1\rangle$ with different spin backgrounds $|\sigma_{j,{\bm \delta}_1}\rangle$ always retain orthogonality in subsequent evolution under $H_0+H_{\rm dip-dip}$. This implies that the excited p-orbital $|\bfr_j+{\bm \delta}_1\rangle$ in $|1\rangle$ is ``distinguishable'' in complete analogy to the situation studied in Sec.~\ref{subsec:distinguishable}, where we considered an array in the quantum optics limit with a single atom selectively driven by an external source. In particular, each orbital configuration $|\bfr_j+{\bm \delta}_1\rangle$ represents an excitation deposited on a selected atom, which can spread inside the array through the propagator $G_{\chi}(\delta)=-(H_0+H_{\rm dip-dip})^{-1}$ (in the rotating frame of the incident light) defined in Sec.~\ref{subsec:distinguishable}. One of the consequences is an effective decay rate $\Gamma_{\rm d}(\delta)=-2\im 1/\chi(0,\delta)$ as seen by the distinguishable excitation, which is depicted by red, wavy arrows in Fig.~\ref{fig:chemistry_optics}b from the state $|1\rangle$. This analogy is manifestly seen once we use the Nakajima-Zwanzig formalism~\cite{reiter_effective_2012} to integrate out the excited states $|\bfr_j+{\bm \delta}_1\rangle$ and the continuum to which they couple, to produce an effective non-Hermitian dynamics on state $|E_{\bfk_{xy}=0}\rangle$. The resulting complex self-energy, encoding the coherent energy shift and decay rate due to this coupling to a continuum, is given by~\cite{reiter_effective_2012} $\Sigma_{t}(\delta)=\bra{E_{\bfk_{xy}=0}} H_t  G_{\chi}(\delta) H_t\ket {E_{\bfk_{xy}=0}}
=-4t_{\rm eff}^2\chi(0,\delta)$. The appearance of the susceptibility $\chi$ defined by a \textit{classical} optics calculation confirms the analogy.

We now consider the opposite limit where $H_t$ dominates the subsequent dynamics of the state $|1\rangle$. Besides returning back to $|E_{\bfk_{xy}=0}\rangle$, $H_t$ connects $|1\rangle$ to an additional orthogonal state $|2\rangle$ characterized by $n=2$ non-trivial hops of the p-orbital relative to its position in the original state $|E_{\bfk_{xy}=0}\rangle$,
\begin{equation} |2\rangle=\frac{1}{2\sqrt{3N}}\sum_j \sum_{{\bm \delta}_{1,2}=\pm d\vhat{x},\pm d\vhat{y}} \left(1-\delta_{{\bm \delta}_1,-{\bm \delta}_2}\right) |\bfr_j+{\bm \delta}_1+{\bm \delta}_2\rangle|\sigma_{j,{\bm \delta}_1,{\bm \delta}_2}\rangle. \label{eq:2}\end{equation}

The corresponding matrix element is $\langle 2|H_t|1\rangle=-\sqrt{3}t_{\rm eff}$. The state $|2\rangle$ has an increased number of nearest neighbors with broken N\'{e}el ordering, with the spin states $|\sigma_{j,{\bm \delta}_1,{\bm \delta}_2}\rangle$ being orthogonal to one another and to the spin states in $|1\rangle$ and $|E_{\bfk_{xy}=0}\rangle$. For a larger number of hops $n>2$, a standard approximation is to assume that spin backgrounds are always distinguishable~\cite{brinkman_single-particle_1970,dagotto_correlated_1994}. Then, the problem reduces to hopping on a Bethe lattice and the matrix elements are $\langle n+1|H_t|n\rangle=-\sqrt{3}t_{\rm eff}$ for $n\geq 1$, as discussed in Appendix~\ref{app:bethe}.

Intuitively, the effect of hopping over the states $|n\rangle$ will dominate the effective dissipation seen by the state $|E_{\bfk_{xy}=0}\rangle$ when 
$t_{\text{eff}}\gg \Gamma_{\rm d}(\delta)$. As shown in \figref{fig:parameter_estimates}, in the relevant range of lattice constants $d\gg a_0$, this regime never occurs when illuminating the system exactly at the resonance $\delta=\omega(0)$. However, hopping to other states $|n\rangle$ can become important for other near-resonant driving frequencies $\delta\neq \omega(0)$. Hopping on the Bethe lattice has been previously solved in Ref.~\cite{mahan_energy_2001}, with the main results summarized in Appendix~\ref{app:bethe}. In particular, one finds that these dynamics contribute an imaginary self-energy to the the excited state $|E_{{\bf k}_{xy}=0}\rangle$, $\Sigma_t(\delta)= -4i t_{\text{eff}}$. This intuitively states that the effective decay rate from $|E_{{\bf k}_{xy}=0}\rangle$ to the continuum of states $|n\rangle$ is proportional to the hopping matrix element itself.

Up to now, we have considered the limits where either $H_t$ or $H_{\rm dip-dip}$ dominates the dynamics from the state $|1\rangle$. To include both effects, we can use the simple, phenomenological formula
\begin{equation} 
\Sigma_t(\delta) =  \dfrac{4t_{\text{eff}}^2\chi(\delta,0) }{i t_{\text{eff}} \chi(\delta,0)-1},\label{eq:Sigma_t} 
\end{equation}
which interpolates between the results obtained in the two limits. This is the main result of Sec.~\ref{subsec:photo_excited_dynamics}, as it reduces all of the chemistry-induced photon-excited electron dynamics to an effective complex self-energy correction to the excited state $|E_{{\bf k}_{xy}=0}\rangle$.


\subsection{Density-density correlations}

\label{subsec:density_correlations}

We now ignore the p-orbital dynamics of $H_t$, and consider just the effect of ground-state density-density correlations under the quantum optics Hamiltonian $H_{\rm QO}$. 
The holon (nucleus with no electron) and doublon (approximately a negatively charged hydrogen ion) have a completely different response to light and in particular do not efficiently couple to light near resonance with the neutral hydrogen transition. At large $d/a_0$, we can thus model the optical response of the holon-doublon pair in the otherwise perfect array as a \textit{classical} array of point dipoles with two consecutive empty sites. The breaking of discrete translational symmetry by these two sites induces light scattering from the incident direction into random ones, effectively leading to an imaginary contribution to the index. The fact that light scattering is sensitive to density-density correlations is well-known in other contexts, for example, forming the foundation for inelastic x-ray scattering spectroscopy~\cite{sturm_dynamic_1993}.

Specifically, we want to quantify the optical properties of an array with a fraction $P_{\rm hd} \ll 1$ of holon-doublon pairs, as defined in \eqrefTemp{eq:hd_population} (i.e. number of pairs over the total number of lattice sites). Assuming that the density of pairs is low enough that the emission from different pairs is uncorrelated, we can proceed in an analogous fashion to Sec.~\ref{subsec:distinguishable}, where we calculated the optical response of an array with a small fraction $P_h$ of random holes. Analogous to Eq.~(\ref{eq:t_holes}), we find that 
\eq{
\label{eq:optical_response_t_r_single_defect}
r(\delta ) \approx  
 \dfrac{i\Gamma(0)/2}{-\delta +\omega(0) -i \Gamma(0)  /2 +  P_{\rm hd} \Sigma _{\text{hd}}(\delta)},\;\;\;\;\;\;\;t(\delta)=1+r(\delta),
}
where we have defined the complex self-energy $\Sigma_{\text{hd}}(\delta)= \sum_{\bm{\hat{\xi}}=\vhat x,\;\vhat y} 1/[\chi(0,\delta)+\chi(d \bm{\hat{\xi}} ,\delta)] $, which is averaged over the two possible orientations $\bm{\hat{\xi}} =\vhat x,\;\vhat y$ of the holon-doublon pairs. In particular, the imaginary component of $\Sigma_{\rm hd}(\delta)$ characterizes an effective dissipation arising from the scattering of normally incident light $\bfk_{xy}=0$ into random other directions. In practice, the~(modest) difference of the above equation as compared to Eq.~(\ref{eq:t_holes}) is that the scattering between two consecutive sites occupied by a holon-doublon pair is correlated, and one cannot simply make the substitution $P_{\rm h}\rightarrow 2P_{\rm hd}$ in Eq.~(\ref{eq:t_holes}). Importantly, though, a holon-doublon pair retains a relatively large resonant cross section to scatter into other directions, close to the value of Eq.~(\ref{eq:Nh}). This can be understood by noticing that the pair strongly scatters into the isoenergetic modes of \figref{fig:distinguishable}b that satisfy $\omega(\bfk_{xy})\approx \omega(0)$ (black dashed line), which are roughly characterized by $|k_x|\sim \text{const}\ll \pi/d$, and $|k_y|\leq \pi/d$. When the holon-doublon pair is oriented along $\vhat{x}$, the relevant Bloch modes cannot resolve the two defects placed at a distance $d$, and the total scattering cross section is then very close to that of a single defect. On the contrary, when the connecting vector between the pair of sites is along $\vhat{y}$, the Bloch modes can resolve the two sites, and the scattering cross section is roughly twice that of a single defect, in agreement with numerical evidence.


\begin{figure}[t]
\centering
\includegraphics[width=0.9\textwidth]{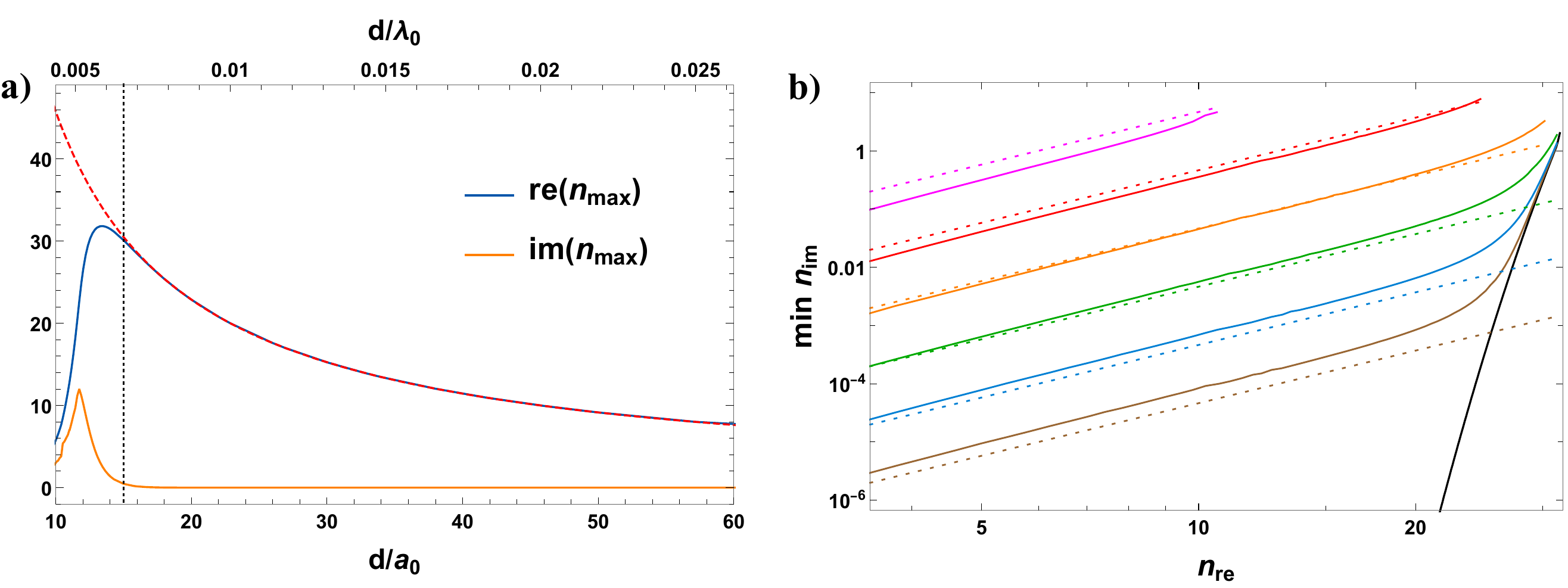}
\caption{Optimal refractive index. a) Approximate calculation of maximum real part of the refractive index (solid, blue line, maximized over the detuning $\delta$), as a function of lattice constant, including the effects of quantum chemistry $\Sigma_{\text{QC}}(\delta)$, without additional phenomenological decay $\Gamma'=0$. The dashed black curve represents the result $n=\lambda_0/(2d_z)$ in the quantum optics limit, while the red curve shows the imaginary part of the index. b) Given any fixed value of the real index $ n_{\text{re}}$, we plot the minimum imaginary part $n_{\text{im}}$, obtainable with proper choices of $d$ and $\delta$ (colored, solid lines). The curves (from brown at the bottom, to magenta on top) refer to increasing values of the additional inelastic losses $\Gamma'=0.01,\;0.1,\;1,\;10,\;10^2,\;10^3$. The black solid line shows the case $\Gamma'=0$, which is purely limited by the intrinsic effects of $\Sigma_{\text{QC}}(\delta)$. The dotted, colored lines represent the asymptotic scaling  $ n_{\text{im}} \sim  [
k_0^3 d_{\text{QC}}^3 \Gamma' /(12 \pi \Gamma_0)] n_{\text{re}} ^3   $, where $d_{\text{QC}}\approx 15 a_0$ approximately represents the lattice constant where the effects of quantum chemistry become relevant (black, dotted vertical line of \figref{fig:index}-a).
}
\label{fig:index}
\end{figure}


\subsection{The limit of refractive index by quantum chemistry}\label{subsec:index_limit}

From the previous subsections, we can assign a total complex self-energy $\Sigma_{\text{QC}}(\delta)=\Sigma_{ \text t}(\delta)+P_{\text{hd}}\Sigma_{\text{hd}}(\delta)$ to the collective mode $|E_{\bfk_{xy}=0}\rangle$ of a 2D array, which includes the effects of both the p-orbital dynamics $\Sigma_{\text t}$ via Eq.~(\ref{eq:Sigma_t}) and the density-density correlations $\Sigma_{\text{hd}}$ via Eq.~(\ref{eq:optical_response_t_r_single_defect}). The frequency-dependent resonance shift $\re \Sigma_{\text{QC}}(\delta)$ and inelastic losses $-2\im \Sigma_{\text{QC}}(\delta)$ alter the linear reflection and transmission coefficients in response to a normally incident field, to $r(\delta)= i\Gamma(0) /[-2\delta+2\omega(0) + 2\Sigma_{\text{QC}}(\delta)-i \Gamma(0) ]$ and $t(\delta)=1+r(\delta)$. A non-zero loss $-2\im \Sigma_{\text{QC}}(\delta)>0$ generally results in a loss of coherently scattered energy $|r|^2+|t|^2<1$. It should also limit the maximum index achievable. This can easily be seen in the limit of large $-2\im \Sigma_{\text{QC}}(\delta)/\Gamma(0)\gg 1$, where $r(\delta)\sim 0$ and $t(\delta)\sim 1$ indicating that the array ceases to respond to light altogether.

The derivation of the refractive index of a 3D lattice, based upon multiple scattering between 2D arrays, follows in a manner analogous to that presented in Sec.~\ref{subsec:indexQO}. In particular, recall that we obtained the dispersion relation $J(k_z)$ of Eq.~(\ref{eq:optical_band_equation}) for a 3D system by diagonalizing the Hamiltonian $H^{\rm 1D}_{{\rm dip-dip},ij}$ of Eq.~(\ref{eq:H1Ddip-dip}) describing field-mediated interactions between planes. Within the limits that we consider quantum chemistry, one can repeat the calculation with the modification of the intra-plane matrix element $H^{\rm 1D}_{{\rm dip-dip},ii}\rightarrow H^{\rm 1D}_{{\rm dip-dip},ii}+\Sigma_{\rm QC}(\delta)$ to include chemistry effects. This modifies the dispersion relation of Eq.~(\ref{eq:optical_band_equation}) into the nonlinear form $J(k_z)\rightarrow J(k_z)+\Sigma_{\rm QC}(J(k_z))$. 

By choosing an aspect ratio of $d_z/d=2.5$, we can ensure that the contribution of the evanescent field to the band (i.e. $J_{\text{ev}}(k_z)$, as defined in~\eqrefTemp{eq:Jev_full_equation}) is negligible in the range of interest $d/a_0\gg 1$, guaranteeing that the modified dispersion relation Eq.~(\ref{eq:optical_band_equation}) is readily invertible (see Appendix \ref{app:invertible_band} for more quantitative details). By defining the complex refractive index as $n(\delta)=k_z(\delta)/k_0$, we obtain
\eq{
\label{eq:index_final_losses}
n(\delta) =\dfrac{1}{k_0 d_z}\arccos\left[\cos(k_0 d_z) +  \dfrac{ \Gamma(0) \sin(k_0 d_z)}{ 2\delta -  2\omega(0)- 2\Sigma_{\text{QC}}(\delta) + i\Gamma'  } \right].
}
To generalize the derivation for subsequent discussions about possible experimental realizations, we have also included a phenomenological inelastic loss term $\Gamma'$ to the self-energy, $\Sigma_{\rm QC}(\delta)\rightarrow \Sigma_{\rm QC}(\delta)-i\Gamma'/2$, which accounts for other effects beyond the quantum chemistry interactions explicitly considered up to now. One can prove that the definition of index leading to Eq.~(\ref{eq:index_final_losses}) correctly describes the optical properties within the framework of classical macroscopic electrodynamics. For example, in Appendix~\ref{sec:index_classic_optics}, we show that the formula of $n(\delta)$ correctly describes the reflection and transmission of a finite-length 3D system when inserted into standard Fresnel coefficient formulas for a dielectric slab, as long as the wavelength of light cannot resolve the atomic positions, i.e. when $k_0d_z \ll 1$ and $|n(\delta) k_0d_z| < 1$.

Eq.~(\ref{eq:index_final_losses}) represents our final formal result, where we are able to transition from the quantum optics to~(weak) quantum chemistry limit, while calculating the refractive index in a manner that still retains non-perturbative multiple scattering of light. In order to appreciate its non-perturbative nature, we can examine the requirements for Eq.~(\ref{eq:index_final_losses}) to reduce to usual perturbative theories of optical response, such as the Drude-Lorentz model. As shown in Appendix~\ref{app:Drude-Lorentz}, this occurs when the inelastic losses due to quantum chemistry become so intense as to strongly suppress the effects of multiple scattering, specifically, when $-2\im \Sigma_{\text{QC}}(\delta)/\Gamma(0)> 1$ and $k_0d_z\ll 1$. This observation helps to qualitatively understand why perturbative theories of optical response work so well when quantum chemistry interactions become strong, as is the case for real solids.

We use \eqrefTemp{eq:index_final_losses} to calculate the complex refractive index first considering $\Gamma'=0$, as a function of the lattice constant $a_0 \ll d \ll \lambda_0$, choosing the detuning $\delta$ which maximizes its real part. 
In the numerical implementation, we must avoid the range of frequencies associated with the bandgap, where there are no propagating modes, i.e. the range of values of $J$ that have no solution for any $k_z$. It can readily be checked that even if the losses are explicitly set to zero~($\im\,\Sigma_{\rm QC}=\Gamma'=0$), within the bandgap region, Eq.~(\ref{eq:index_final_losses}) would predict a complex index, incorrectly suggesting a lossy medium. We emphasize that this issue is simply associated with how to define a proper macroscopic index in the bandgap regime, whereas the microscopic dispersion relation $J(k_z)$ remains correct.
 
The results of \eqrefTemp{eq:index_final_losses} for $\Gamma'=0$ are shown in \figref{fig:index}-a, where the blue line shows the maximum real part of the index, while the orange line represents the associated imaginary part (i.e. at the same frequency $\delta$). The red, dashed line shows the ideal quantum optics scaling of $n_{\rm max}=\lambda_0/2d_z$ obtained in \eqrefTemp{eq:fra_max_index_invertible}. One can see that the model predicts a possible real part of the index as large as $\max n_{\rm re}\approx 30$ around $d \approx 15 a_0$, accompanied by a small imaginary part describing losses $n_{\rm im}\lesssim 1$, for an optimal lattice constant. As one further decreases the lattice constant, one first sees a decrease in the real part of the index and an increase in the imaginary part, followed by a decrease in both, even as the effects of quantum chemistry continuously increase, as characterized by $\Sigma_{\rm QC}(\delta)$. This reflects our earlier observation that a huge inelastic loss rate should make an individual 2D layer increasingly transparent.

Rather than focus on how large the real part of the index can be, a more relevant question might be how small the loss can be, $\min\,n_{\rm im}$, given a target value of the real part of the index $n_{\rm re}$. In \figref{fig:index}b~(solid black curve), we calculate the minimum loss for a target $n_{\rm re}$, optimizing over the lattice constant $d$ and detuning $\delta$. We next consider how robust our ultrahigh index, low loss material is to hypothetical additional dissipation rates $\Gamma'>0$ beyond the specific quantum chemistry interactions that we considered. In \figref{fig:index}b we repeat the same analysis, but including in Eq.~(\ref{eq:index_final_losses}) a range of values $\Gamma'/\Gamma_0=0.01,\;0.1,\;1,\;10^2,\;10^3$ (from the bottom, solid brown curve to the top, solid magenta curve). The dotted lines represent the scaling $ \min\,n_{\rm im} \sim  [
k_0^3 d_{\text{QC}}^3 \Gamma' /(12 \pi \Gamma_0)] n_{\rm re}^3$, an approximate result derived in Appendix \ref{app:im_re_index}, and which is valid as long as $n_{\rm im}\ll n_{\rm re}$. There, $d_{\text{QC}}\approx 15 a_0$ roughly represents the lattice constant where quantum chemistry starts to play a major role (as shown by the dotted, vertical line in \figref{fig:index}a).

From the above discussions, it is clear that in principle one possible approach to achieve high-index materials is to realize high-density arrays of well-positioned, sufficiently homogeneous quantum emitters~\cite{khurgin_expanding_2022}. The maximum index would be achieved at a distance between emitters right before the electronic orbital wave functions between nearest neighbor emitters begins to appreciably overlap. Although we know of no specific platform that immediately allows for an ultrahigh index, we note that there has been steady progress to deterministically position emitters, such as by self-organization \cite{raino_superfluorescence_2018} or ion beam implantation~\cite{schroder_scalable_2017}. We also note that in principle, quantum emitters already exist with sufficiently small values of $\Gamma'$~(where we allow $\Gamma'$ to incorporate non-radiative decay, additional undesired radiative decay paths, dephasing, and inhomogeneous broadening) that an ultrahigh index might be possible, if they could be arranged into arrays. For example, single color centers in diamond (such as silicon-vacancy centers) exhibit inelastic rates as low as $\Gamma' \sim \Gamma_0$~\cite{rogers_electronic_2014,schroder_scalable_2017}, and inhomogeneous broadening levels at low temperatures in the range of $\Gamma'/\Gamma_0 \sim 10$-$100$~\cite{rogers_multiple_2014,evans_narrow-linewidth_2016,schroder_scalable_2017}. Single quantum dots can offer almost lifetime-limited linewidths with $\Gamma'\ll \Gamma_0$~\cite{kuhlmann_transform-limited_2015,pedersen_near_2020}, although some technological improvement is still required to reduce the amount of inhomogeneous broadening in ensembles. Separately, since the key underlying ingredient for high index is a near-ideal single-mode response of a single 2D layer, 2D materials supporting excitonic resonances could also be a suitable platform. In particular, 2D transition metal dichalcogenides have been observed to exhibit nearly perfect reflection on resonance~\cite{zeytinoglu_atomically_2017,scuri_large_2018,back_realization_2018}, due to the high radiative efficiency of excitons in such systems. If such individual layers with sufficiently low loss could be stacked with controllable spacings between layers~\cite{novoselov_2D_2016}, an ultrahigh index should exist until quantum chemistry between layers becomes appreciable and the index reduces back to the value found in bulk 3D material.

\section{Conclusions and outlook}\label{sec:outlook}

In summary, we have shown that the magnitudes of refractive indices observed in known optical materials likely does not reflect a fundamental limit, and an ultrahigh index, low-loss material should be allowed by the laws of nature. Our analysis also suggests why an answer to the problem surrounding the limits of refractive index has been elusive, as the answer seemingly requires one to understand the nature of non-perturbative multiple light scattering over a broad range of densities that spans across the quantum optics and quantum chemistry limits. Our work will hopefully stimulate new efforts to identify, design, and fabricate ultrahigh index materials.

While our current analysis focused on a specific model in which the limits to index arise due to electronic density-density correlations and dynamics of excited electrons, it would be interesting in future work to examine other general material models. For example, are there paradigms in which the mechanisms discussed here can be strongly suppressed, leading to higher indices? In order to better answer such questions, and also to aid in the search or possible design of ultrahigh-index materials, it might also be desirable to develop more general frameworks for the calculation of optical response in the regime of non-perturbative scattering, and which ideally might be integrated with state-of-the-art computational quantum chemistry. One promising approach might be to generalize the electromagnetic Green's function based methods to many-body condensed matter settings.

Finally, while our work specifically focused on the question of linear refractive index, it more generally suggests that there is a broad range of material densities where other important optical properties might have surprising behavior, due to strong multiple scattering. As one example, it would be interesting to develop similar theories for the limits of nonlinear optical response, and to address whether there exist mechanisms to enhance the nonlinear response beyond that of known materials.


\begin{acknowledgments}
The authors would like to thank A. Browaeys, L. Gagliardi, and M. Hafezi for enlightening discussions. F.A. acknowledges support from the ICFOstepstone - PhD Programme funded by the European Union’s Horizon 2020 research and innovation programme under the Marie Skłodowska-Curie grant agreement No 713729. D.E.C. acknowledges support from the European Union’s Horizon 2020 research and innovation programme, under European Research Council grant agreement No 101002107 (NEWSPIN) and FET-Open grant agreement No 899275 (DAALI); the Government of Spain (Europa Excelencia program EUR2020-112155 and Severo Ochoa Grant CEX2019-000910-S [MCIN/AEI/10.13039/501100011033]); Generalitat de Catalunya (CERCA program and AGAUR Project No. 2021 SGR 01442); Fundació Cellex; and Fundació Mir-Puig.
\end{acknowledgments}

\cleardoublepage
\appendix

\section{Invertibility of the optical band structure}
\label{app:invertible_band}
\begin{figure}[b!]
\centering
\includegraphics[width=\textwidth]{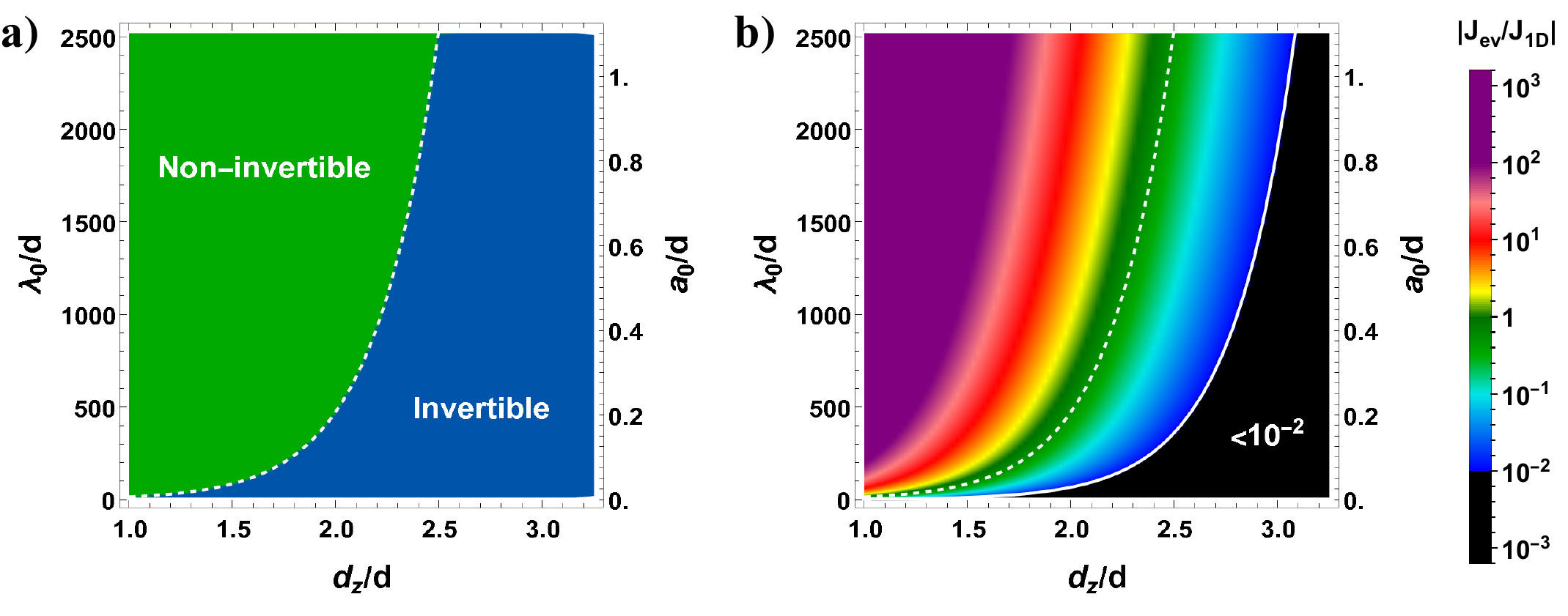}
\caption{\textbf{Contribution of the evanescent fields to the band dispersion of a 3D lattice.} a) 
Regimes where the band is either invertible (blue region) or not (green region), as a function of the aspect ratio $d_{z}/d $ and the longitudinal lattice constant $d_{z}/\lambda_0$. The data are calculated by fully numerically computing the dispersion relation $J(k_z)$ and explicitly looking for local maxima at some $|k_z|<\pi/d_z$. The white, dashed line represents the analytic bound of Eq.~\eqref{eq:invertible_analytic_bound}.
b) Ratio between the evanescent and radiative contributions to the band structure, i.e. $|J_{\text{ev}}/J_{\text{1D}}|$, calculated at $k_z=\pi/d_z$. The dashed, white line is the same invertibility boundary as before, which is equivalent to $|J_{\text{ev}}/J_{\text{1D}}|\lesssim 1/2$. The solid, white line shows the analytic prediction for the threshold where $|J_{\text{ev}}/J_{\text{1D}}|\leq 10^{-2}$. In both plots the value of $J_{\text{ev}}$ is computed numerically from its exact formula of Eq.~\eqref{eq:Jev_full_equation}.
}
\label{fig:invertible_band_regime}
\end{figure}

In this section, we discuss the effect of the evanescent interaction between atomic layers in the quantum optics regime, which can make the optical band structure non-invertible. We start from the band structure of the 3D system $J(k_z)$, as described in Eq.~\eqref{eq:optical_band_equation} and Eq.~\eqref{eq:Jev_full_equation}. We are interested in the limit $d_z\ll \lambda_0$ and $d_z\geq d$. By Taylor expanding in the ratio $|\cos\left(k_z d_z\right) /\cosh\left(   \left|{\bf g}_{mn}  \right| d_{z} \right)|\ll 1$, one can simplify the evanescent contribution to the dispersion relation to $J_{\text{ev}}(k_z) \approx (\lambda_0/d_z) \left[ -A (  d_z/d  )  + B (  d_z/d  )\;\cos(k_z d_z)\right]$,
where we define the coefficients
\eq{
\label{eq:band_A_and_B_analytic}
A(d_z/d)=\left(\dfrac{d_z}{d}\right)\Sum_{\substack{m \in \mathbb Z\\ n \in \mathbb Z\\(m,n)\neq (0,0)}} \dfrac{  m ^2}{\sqrt{ m ^2+ n ^2 }}\left[ 1- \tanh\left(2\pi (d_z/d) \sqrt{ m^2+ n ^2 }\right)  \right]
,\\\\
B(d_z/d)=\left(\dfrac{d_z}{d}\right)\Sum_{\substack{m \in \mathbb Z\\ n \in \mathbb Z\\(m,n)\neq (0,0)}} \dfrac{  m ^2}{\sqrt{ m ^2+ n ^2 }}\left[\dfrac{\tanh\left(2\pi (d_z/d) \sqrt{ m^2+ n ^2 }\right) }{ \cosh\left(2\pi (d_z/d) \sqrt{ m^2+ n ^2 }\right)  }\right]
.
}
which only depend on the aspect-ratio of the lattice $d_z/d$. This allows to easily calculate the properties of the band in an analytic fashion. The presence of a local maximum in $J_{\rm ev}$ around $|k_z|\lesssim \pi/d_z$ is responsible for the non-invertible behavior of the band. In the limit $k_0d_z\ll 1$, the condition for this local maximum to exist becomes
\eq{
\label{eq:invertible_analytic_bound}
\dfrac{d}{\lambda_0}<\dfrac{d}{d_z}\sqrt{\dfrac{2B(d_z/d)}{\pi}}
\approx 2\sqrt{\dfrac{2d}{\pi d_z}}e^{-\pi d_z / d},
}
which defines the regime where the band is non-invertible. This threshold is represented by the white, dashed line of \figref{fig:invertible_band_regime}-a, where we illustrate the condition as a function of $\lambda_0/d$ and $d_z/d$. In the same figure, we perform an exact numerical calculation of the band structure equation~(\ref{eq:optical_band_equation}), and indicate with blue and green the regions of parameter space where the band structure is invertible and non-invertible, respectively. We see that the approximate condition of Eq.~(\ref{eq:invertible_analytic_bound}) agrees well.

The condition above describes when $J_{\text{ev}}(k_z)$ is so strong to radically alter the band structure, making it non-invertible. Here, we want to quantify when the evanescent field negligibly contributes to the overall band structure. We thus calculate the ratio between the evanescent $J_{\text{ev}}(k_z)$ and the radiative $J_{\text{1D}}(k_z)=\sin(k_0 d_z)/[\cos(k_z d_z)-\cos(k_0 d_z)]$ contributions. This quantity in principle depends on the wavevector $k_z$, so we consider the maximum value $\max_{k_z} |J_{\text{ev}}/J_{\text{1D}}|$. When $d_z\ll \lambda_0$, this ratio is maximized by $k_z=\pi/d_z$ and it reads $ \max |J_{\text{ev}}/J_{\text{1D}}|\approx  \left( \lambda_0/d_z\right)^2  B(d_z/d)/\pi$. Comparing this with Eq.~(\ref{eq:invertible_analytic_bound}), one can deduce that the band becomes non-invertible if $|J_{\text{ev}}/J_{\text{1D}}|\gtrsim 1/2$. In \figref{fig:invertible_band_regime}b, we represent the value of this maximum ratio as a function of the aspect ratio $d_z/d$ and lattice constant $d$. The black region represents the regime where $|J_{\text{ev}}/J_{\text{1D}}|\geq 10^{-2}$, which allows us to ignore the evanescent contribution. For our choice of the aspect ratio $d_z/d=2.5$, this is true as long as $d/a_0\gtrsim 6$.

\begin{figure}[t]
\centering
\includegraphics[width=0.5\textwidth]{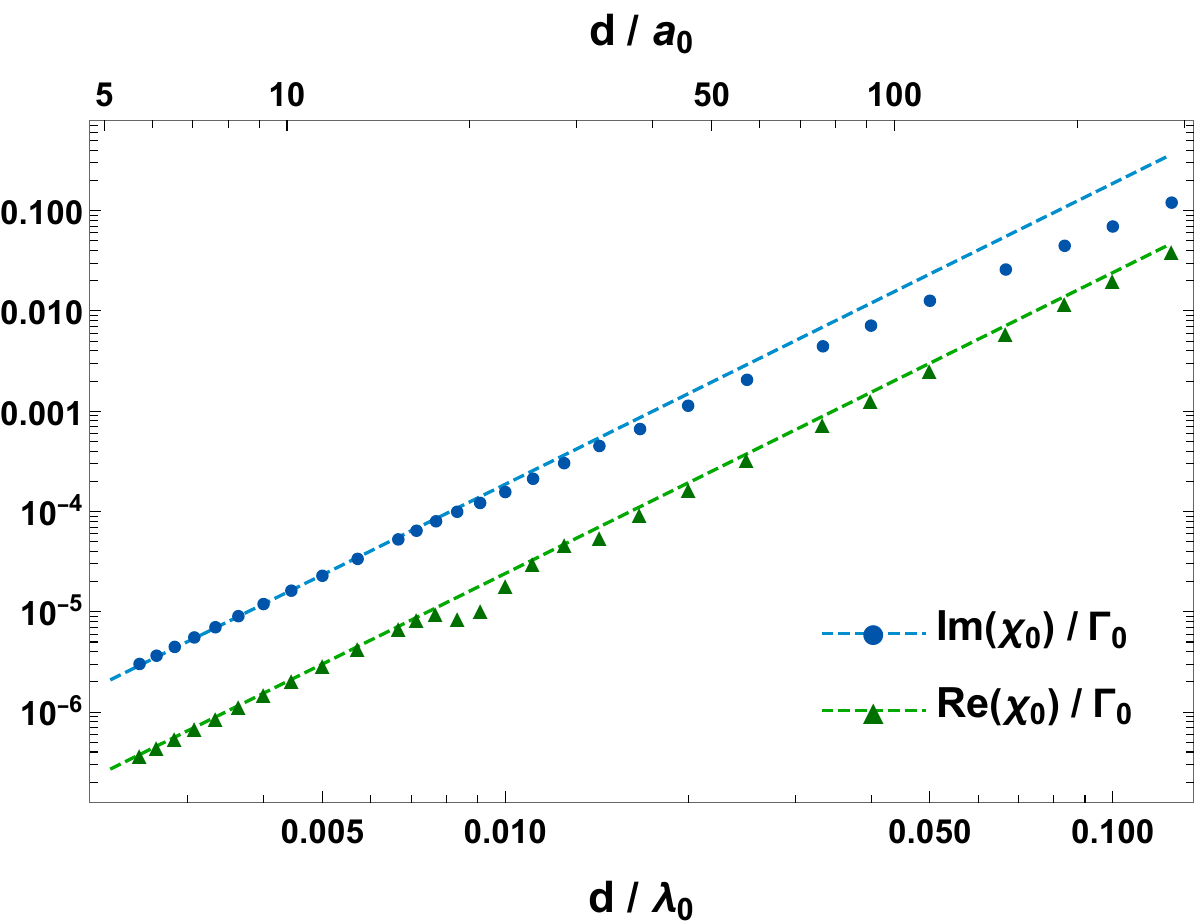}
\caption{
Plot of the resonant susceptibility of the selectively driven atom $\chi_0=\chi(0,\delta=\omega(0))$ versus $d/\lambda_0$~(lattice constant normalized by resonant wavelength). The blue points (green triangles) represent, in log-log scale, the value of its imaginary (real) part, while the dashed blue and green lines show the asymptotic scalings $\im\chi_0 / \Gamma_0 \sim  187 (d/\lambda_0)^3$ and $\re\chi_0 / \Gamma_0 \sim  24 (d/\lambda_0)^3$. On the top $y$-axis, we also label the corresponding ratio of lattice constant to Bohr radius $d/a_0$, assuming that the resonant wavelength $\lambda_0$ is taken to be that of the hydrogen atom.
}
\label{fig:chi_re_im_FINAL}
\end{figure}

\section{Susceptibility of a distinguishable atom}
\label{app:distinguishable_atom_gamma_d}
In this appendix, we detail how to numerically calculate the susceptibility $\chi(\vecTemp r_j-\vecTemp r_h,\delta)=c_j(\delta)/\Omega_h$ from the steady-state solutions $c_j(\delta)$ of an infinite 2D array of lattice constant $d$, where one atom at position $\vecTemp r_h$ is selectively driven by a near-resonant Rabi frequency $\Omega_h$, detuned by a factor $\delta=\omega_L-\omega_0$. We recall that the atomic wave function is given by $|\psi_{\rm 2D}(t)\rangle=c_G(t)|G\rangle+\sum_j c_j b_{pj}^{\dagger}b_{sj}|G\rangle$, and the other atoms $\vecTemp r_j\neq \vecTemp r_h$ can still be excited via dipole-dipole interactions with the driven atom, via \eqrefTemp{eq:3DQO_Hamiltonian}. 

We numerically simulate a finite 2D square array of lateral size $2l$ and lattice constant $d$. For simplicity, the selectively driven atom is placed at the center $\vecTemp r_h=0$. To mimic the infinite size of the array, we introduce an additional non-Hermitian term to the Hamiltonian, $H_{\rm nr}=-(i/2)\sum_{j}\Gamma'_{\rm cut-off}(\bfr_j)b_{pj}^{\dagger}b_{pj}$. This term adds an extra non-radiative decay to the excited p-orbitals, which has a smooth position dependence $\Gamma'_{\text{cut-off}}(R=\sqrt{x_j^2+y_j^2})$, and its purpose is to smoothly dissipate energy that propagates outward along the array towards the boundaries in order to suppress multiple reflections at the boundaries. Specifically, we define 
\eq{
\label{eq:cut_off_numerics}
\Gamma'_{\text{cut-off}}(R)=\left\{\arr{ll}{
0 & \text{  if   } R\leq R_{\text{cut-off}}\\\\
3\Gamma(0)\left(\dfrac{R-R_{\text{cut-off}}}{R_{\text{cut-off}}/ 2 }\right)^2  & \text{  if   } R> R_{\text{cut-off}} 
}\right..
}
On top of that, we fix $R\leq l=(3/2)R_{\text{cut-off}} $. This way, $\Gamma_{\rm cut-off}'(R_{\rm cut-off})=0$ and  $\Gamma'_{\text{cut-off}}(l)=3\Gamma(0)$ at the boundaries $R=l$. The fact that the dissipation is zero around the position of the driven atom $\bfr_h=0$, and smoothly growing for increasing $R$, makes the numerically calculated susceptibility essentially independent of the details of the added decay.

Due to the finite size of the system, we are effectively computing the optical response by discretely sampling Bloch wavevectors, rather than accounting for the full continuum. The smallest wavevector that we are implicitly considering can be roughly estimated by $|\bfk_{xy}^{\rm min}|\approx \pi/R_{\rm cut-off}$. We thus impose in our numerics that $R_{\rm cut-off}\geq \lambda_0/2$, aiming to well capture at least those modes outside the light cone $|\bfk_{xy}|=k_0$. For lattice constants as small as $d\simeq \lambda_0/400$, this condition implies atomic numbers as large as $N\approx 4\times 10^5$, which represents the maximal size that we can simulate. 
For larger lattice constants $d \gtrsim \lambda_0/150$, however, our numerics can tolerate larger systems, and in that case we impose $N\geq 4\times 10^4$, to reduce the extent of our numerical approximations.

Our results are exemplified in Fig.~\ref{fig:chi_re_im_FINAL}, where we numerically calculate the resonant susceptibility of the selectively driven atom $\chi_0=\chi(0,\delta=\omega(0))$, as a function of lattice constant $d$. Specifically, the blue points and the green triangles show the imaginary and real part of $\chi_0$, along with their asymptotic values $\im\chi_0 / \Gamma_0 \sim  187 (d/\lambda_0)^3$ (dashed blue line) and $\re\chi_0 / \Gamma_0 \sim  24 (d/\lambda_0)^3$ (dashed blue line), which confirm the scaling $\chi_0  \sim   \Gamma_0  (d/\lambda_0)^3$ for small $d$.


\section{Hopping on a Bethe lattice}
\label{app:bethe}


\begin{figure}[t]
\centering
\includegraphics[width=0.45\textwidth]{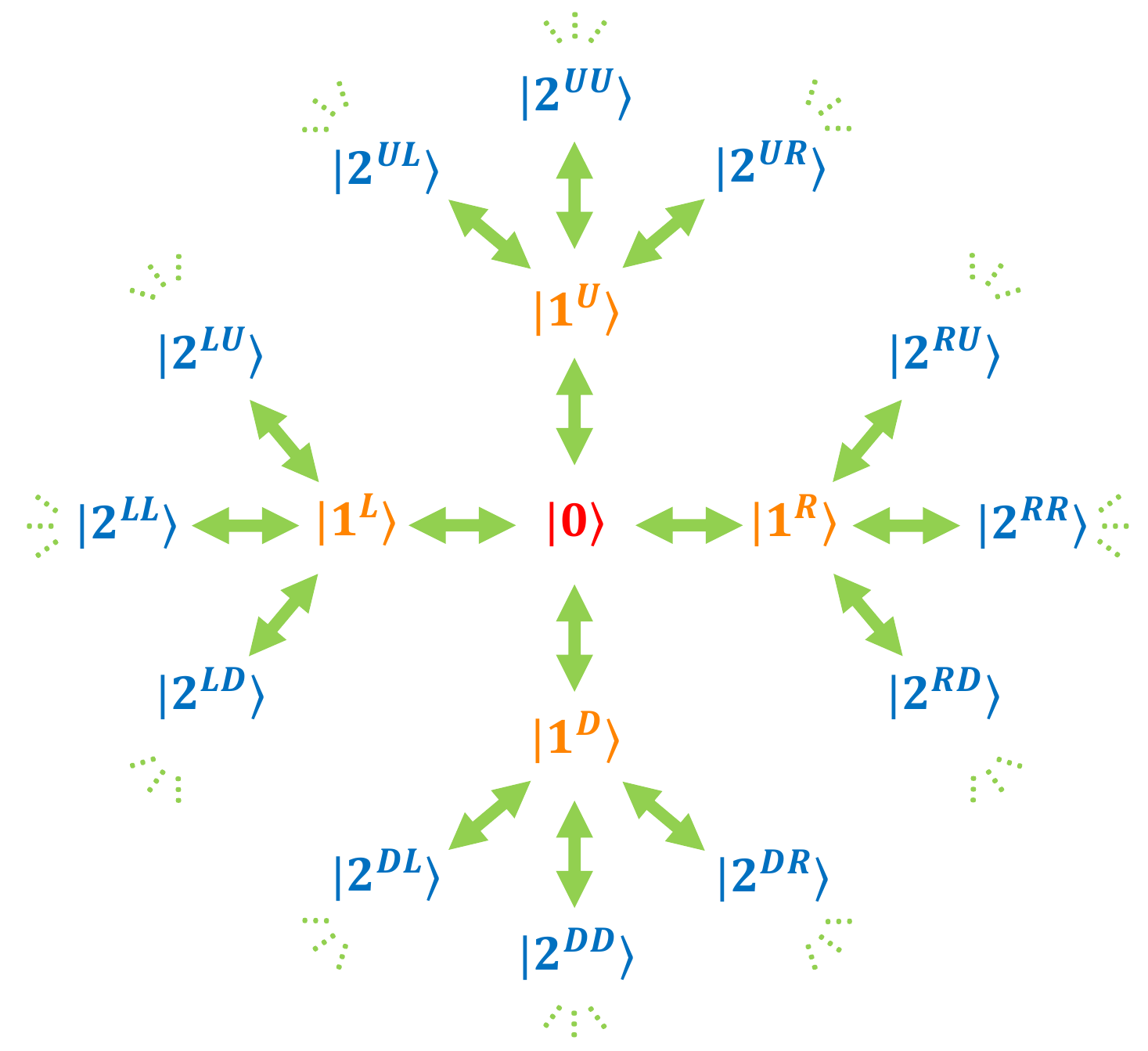}
\caption{\textbf{Behavior of the hopping dynamics due to $H_t$, depicted as a Bethe lattice.} Evolution of the initial excited p-orbital $\ket{0}$, to states $|n^{ab...}\rangle$. Here, the non-negative integer $n$ denotes the number of hops, while $a,b,...\in\{U,D,L,R\}$ denotes the direction~(up, down, left, right) of each hop.
}
\label{fig:bethe}
\end{figure}

In this section, we derive the contribution to the self-energy $\Sigma_t(\delta)\approx -4i t_{\text{eff}}$ of the excited state $|E_{\bfk_{xy}=0}\rangle$, which arises from the dynamics of the p-orbital hopping $H_t$. As can be seen from Eqs.~(\ref{eq:1}) and~(\ref{eq:2}) in the main text, the states involved in up to $n=2$ hops are orthogonal due to their different spin backgrounds, and can be labeled according to the original position~(when $n=0$) $\bfr_j$ of the p-orbital, the four possible moves ${\bm \delta}_1=\pm d\vhat{x}, \pm d\vhat{y}$ to a nearest neighbor at $n=1$, and the three possible moves~(besides returning to $\bfr_j$) ${\bm \delta}_2\neq {\bm \delta}_1$ at $n=2$. As the dynamics under $H_t$ is the same for each $\bfr_j$ up to translation, in what follows we will forget about this label and simply denote the initial position as $|0\rangle$. For visualization, as in Fig.~\ref{fig:bethe}, we will also switch to the labels $U,D,L,R$~(up, down, left, right) for the possible values of ${\bm \delta}_i$. Fig.~\ref{fig:bethe} thus shows how $H_t$ has matrix elements~(green arrows) between the initial state $|0\rangle$ and the superposition state $|1\rangle=(1/2)(|1^U\rangle+|1^D\rangle+|1^L\rangle+|1^R\rangle)$ following $n=1$ hops, and how $|1\rangle$ is connected in turn by $H_t$ to the various configurations $|2^{ab}\rangle$~(with $a,b\in \{U,D,L,R\}$) comprising the state $|2\rangle$. The corresponding matrix elements are $\langle 1|H_t|0\rangle=-2t_{\rm eff}$ and $\langle 2|H_t|1\rangle=-\sqrt{3}t_{\rm eff}$.

While this description up to $n=2$ is exact, a standard approximation for larger $n$ is to assume that the nature of the hopping from $|1\rangle$ to $|2\rangle$ generalizes to any $|n\rangle$ to $|n+1\rangle$~\cite{brinkman_single-particle_1970,dagotto_correlated_1994}. In particular, one assumes that $H_t$ connects a particular configuration $|n^{ab...}\rangle$ to three possible new configurations $|(n+1)^{ab...}\rangle$, and furthermore that all possible generated states have orthogonal spin backgrounds, i.e. the configurations satisfy the orthogonality condition $\langle m^{ab\cdots}|n^{a'b'\cdots}\rangle=\delta_{m,n}\delta_{a,a'}\delta_{b,b'}\delta_{\cdots}$. This makes the problem equivalent to hopping on a so-called Bethe lattice~\cite{mahan_energy_2001}.

Within this approximation, the Hamiltonian $H_t$ takes the form
\begin{equation} H_t\approx H_{\text{Bethe}}\equiv  -2t_{\rm eff}\ket{1}\bra{0}-\sqrt{3}t_{\rm eff} \sum_{n>0}\ket{n+1}\bra{n}+h.c.,\end{equation}
where we define the normalized states $\ket{n}=\sum_{ab\dots}\ket{n^{ab\dots}}/(2\sqrt{3^{n-1}})$, which are an equal superposition of all possible configurations at a given $n$. 
The eigenenergies and eigenstates of $H_{\rm Bethe}$ can be written in the form  
\eq{
E(\theta) = -2\sqrt 3 t_{\text{eff}}\cos(\theta), \\\\
\ket{\psi(\theta) } = \sqrt{\dfrac{2}{\pi}} \left[
 \dfrac{\sqrt 3}{2}\sin( \gamma) \ket{0} + \Sum_{n=1}^\infty \sin(n\theta +\gamma)\ket {n}
\right],
}
where $\tan(\gamma) = 2 \tan(\theta)$. The density of states can be also calculated, obtaining the value \cite{brinkman_single-particle_1970,mahan_energy_2001}
\eq{
\rho(E) = \dfrac{2}{\pi t_{\text{eff}}}\dfrac{\sqrt{24 -(E/t_{\text{eff}})^2}}{16 - (E/t_{\text{eff}})^2 }= -\dfrac{3}{2\pi} \sin^2(\gamma)\dfrac{d\theta}{dE}.
}
Starting from the quasi-bound state $\ket 0$ initially, the decay rate $\Gamma_{\text{Bethe}}$ out of this state into the continuum of the band can be estimated by Fermi's golden rule: 

\eq{
\dfrac{\Gamma_{\text{Bethe}}}{t_{\text{eff}}} =  \dfrac{2\pi}{t_{\text{eff}}} \Int_{-2\sqrt{3}t_{\text{eff}}}^{2\sqrt{3}t_{\text{eff}}} dE\; \rho(E) \left|\bra{\psi(E)} H_{\text{Bethe}}\ket{0}\right|^2 
= 
\dfrac{24  }{\pi }\Int_{0}^\pi d\theta\; \sin^2(\gamma)\sin^2(\theta+\gamma) \\\\
=\dfrac{864}{\pi}\Int_0^1 du\; \dfrac{\sqrt{u^3(1-u)}}{(1+3u)^2}
= 8.
}
This allows us to define the contribution to the self-energy of the state $|E_{\bfk_{xy}=0}\rangle$ as $\Sigma_t(\delta)\approx -4it_{\text{eff}}$.

\section{Fresnel equations of classical optics}
\label{sec:index_classic_optics}

In the main text, we calculated the refractive index~(\ref{eq:index_final_losses}) for a 3D lattice via a Bloch band structure calculation~(with the phenomenological loss term $\Gamma'=0$ set to zero for this discussion). In this appendix, we prove that this definition of index correctly reproduces various predictive properties of optical response, within classical macroscopic electrodynamics. In particular, we on one hand will calculate by microscopic approaches the reflection and transmission through a finite-length 3D system, and on the other hand see that this agrees with the standard Fresnel equations for a dielectric slab.

First, we recall that from the standpoint of classical macroscopic electrodynamics, given a dielectric slab of length $L$ and (complex) refractive index $n(\delta)$, the Fresnel equations predict
\eq{
\label{eq:app_fresnel_t_r}
t_{\text{Fr}}(n) = \dfrac{4 n  e^{i n  k_0 L}}{(1+n )^2-e^{2i n  k_0 L}(n -1)^2},\;\;\;\;\;\;\;\;\;\;
r_{\text{Fr}}(n) = \dfrac{ (n ^2-1)\left(e^{2i n  k_0 L}-1\right)}{(1+n )^2-e^{2i n  k_0 L}(n -1)^2}.
}
We now consider a 3D crystal composed of a number $M$ of 2D atomic arrays, separated by the distance $d_z$ and illuminated at normal incidence. In the regime where the evanescent field can be neglected, each 2D array is characterized by the reflection and transmission coefficients $r(\delta)= i\Gamma(0) /[-2\delta+2\omega(0)+2\Sigma_{\text{QC}}(\delta)-i \Gamma(0)]$ and $t(\delta)=1+r(\delta)$. 
To our analysis, it is convenient to write $r(\delta)$ and $t(\delta)$ as functions of the index $n(\delta)=k_{ z}(\delta)/k_0$. This is accomplished by using Eq.~\ref{eq:index_final_losses} to replace the dependence on $\delta$ with that on $n$, thus obtaining the functions $r(n)$ and $t(n)$. 
The multiple scattering problem through multiple layers reduces to a 1D problem that can be efficiently and exactly solved by the transfer-matrix formalism~\cite{deutsch_photonic_1995}, yielding
\eq{
\label{eq:app_t_r_tot_Cheby}
t_{M}(n)=\dfrac{e^{i k_0 d_z} t(n) }{u_M(n)-e^{i k_0 d_z} t(n) u_{M-1}(n)},\;\;\;\;\;\;\;\;\;\;
r_{M}(n)=\dfrac{e^{2i k_0 d_z} r(n ) u_M(n) }{ u_M(n)-e^{i k_0 d_z} t(n) u_{M-1}(n)}
,
}
where we define the function $u_M(n) = \sin(M n k_0 d_z)/\sin(n k_0 d_z)$. 
Starting from Eq.~\ref{eq:app_t_r_tot_Cheby}, we define the total length $L$ and replace the number of layers with $M\to L/d_z$. Eventually, we can expand the resulting expressions as a Taylor series in $k_0 d_z\ll 1$, with the supplementary assumption that $|n(\delta) k_0 d_z|< 1$, but also keeping in mind that $k_0L$ can be arbitrarily large. By performing such Taylor expansion, one recovers the Fresnel predictions of Eq.~\ref{eq:app_fresnel_t_r}, at the zeroth order in $k_0d_z\ll 1$.


\section{Minimum losses and maximal real index}
\label{app:im_re_index}
In this appendix, we prove the scaling of the minimum imaginary part of the index, given a target real part (optimizing over $\delta$ and $d$), as a function of $\Gamma'$. Starting from \eqrefTemp{eq:index_final_losses}, one can derive the equation 
\eq{
\label{eq:appendix_min_im_start}
2\tilde \delta + \dfrac{\tilde \gamma' \cos(k_0d_z)}{\sin(k_0 d_z n_{\text{re}}) \sinh(k_0 d_z n_{\text{im}})} = \tilde \gamma' \cot(k_0 d_z n_{\text{re}})\coth(k_0 d_z n_{\text{im}}),
}
where $\tilde \delta =  [\delta-\omega(0)-\re \Sigma_{\text{QC}}(\delta)]/\Gamma(0) $ and $\tilde \gamma' = [\Gamma' - 2 \im \Sigma_{\text{QC}}(\delta)]/\Gamma(0)$. We are interested in the regime of low losses $k_0 d_z n_{\text{im}}\ll 1$, in which we can solve \eqrefTemp{eq:appendix_min_im_start}, obtaining 
$
n_{\text{im}} \approx    \tilde \gamma'   [ \cos(k_0 d_z n_{\text{re}}) - \cos (k_0 d_z) ] /[2 k_0 d_z \tilde \delta\sin(k_0 d_z n_{\text{re}}) ]
$
Looking at the numerical optimization of $\min n_{\text{im}}$, one can observe that the minimal losses are always obtained when the lattice constant is roughly fixed at the edge of quantum chemistry, i.e. $d\approx d_{\text{QC}}\approx 15 a_0$, which indeed minimizes the parameter $\tilde \gamma'$ in \eqrefTemp{eq:index_final_losses}. Given $d=d_{\text{QC}}$, the value of $n_{\text{re}}$ is varied by changing $\delta$ on a fixed curve. As we are interested in the regime of low losses $n_{\text{im}}\ll 1$ and negligible quantum chemistry, we can then approximate $\tilde \delta$ with the band structure at $d=d_{\text{QC}}$ (i.e. its analytic solution for a lossless system), reading
\eq{
\label{eq:optical_band_equation_app}
\tilde \delta \approx   
\left( \dfrac{1}{2}\right) \dfrac{\sin(k_0 d_z)}{\cos( n_{\text{re}}   k_0 d_z)-\cos( k_0 d_z)} +\mathcal O\left(\dfrac{n_{\text{im}}}{n_{\text{re}}}\right)^2,
}
which proves to be a good approximation as long as $ n_{\text{im}}\ll n_{\text{re}}$. Then, after expanding for $k_0 d_z\ll 1$ (which is valid only as long as $n_{\text{re}}k_0d_z<1$), one obtains
\eq{
\label{eq:im_n_scaling}
n_{\text{im}} \approx 
\dfrac{\Gamma'}{\Gamma_0}
\left(\dfrac{d_z}{d}\right) 
\left(\dfrac{ k_0^3 d_{\text{QC}}^3}{12 \pi}\right)
\left(\dfrac{1}{n_{\text{re}}} - 2  n_{\text{re}}  +   n_{\text{re}} ^3  \right)  ,
}
where we use the fact that $d=d_{\text{QC}}\approx 15 a_0$ (i.e. before quantum chemistry) to approximate $\tilde \gamma' \approx \Gamma'/\Gamma(0)$.

\section{Recovering Drude-Lorentz due to quantum chemistry}
\label{app:Drude-Lorentz}
At high densities, the losses introduced by quantum chemistry strongly suppress the effects of multiple scattering, preventing the appearance of ultra-high indices. In that regime, one expects that usual mean-field theories (such as Drude-Lorentz) can well describe the physical phenomena. Specifically, we are interested in a system which exhibits one single, dominant resonance $\omega_{\text{res}}$, and which is illuminated by near-resonant light with $|\omega_{\text L}-\omega_{\text{res}}|\ll \omega_{\text{res}}$. In this limit, the Drude-Lorentz model predicts the index
\eq{
\label{eq:n_drude_lorentz}
n=\sqrt{1+\dfrac{f_{\rm res}\omega_{\text P}^2}{\omega_{\text{res}}^2 - \omega_{\text L}^2 - i \gamma'  \omega_{\text L}}}\approx 
\sqrt{1-\dfrac{f_{\rm res}\omega_{\text P}^2/\omega_{\text{res}}}{2(\omega_{\text L}-\omega_{\text{res}}) + i \gamma'  }}
,}
where $\omega_{\text P}$ and $\gamma' $ are respectively the plasma frequency and the damping rate, while $f_{\rm res}$ is the \textit{so-called} oscillator strength which depends on the choice of the resonant transition. Here, we show that the refractive index of an atomic lattice, as predicted by \eqrefTemp{eq:index_final_losses}, reduces to \eqrefTemp{eq:n_drude_lorentz} at high densities, due to the losses induced by quantum chemistry. To this aim, we first re-write \eqrefTemp{eq:index_final_losses} as
\eq{
\cos(k_0 d_z n(\delta)) =  \cos(k_0 d_z) +  \dfrac{ \Gamma(0) \sin(k_0 d_z)}{ 2\delta -  2\omega(0)- 2\Sigma_{\text{QC}}(\delta) + i\Gamma'  }  .
}

At low lattice constants, we can expand this equation up to the second order in $k_0d_z\ll 1$. To do so, one needs to fulfill the condition $|k_0 d_z n(\delta)|<1$, which is guaranteed by the suppression of multiple scattering induced by $-2\im \Sigma_{\text{QC}}(\delta)\gtrsim \Gamma(0)>\Gamma_0$ (this also guarantees the closure of the optical bandgap). This procedure permits to directly recover \eqrefTemp{eq:n_drude_lorentz}, by defining $\omega_{\text{res}}=\omega_0+\omega(0)+\re\Sigma_{\text{QC}}(\delta)$, $\gamma' =\left.\Gamma'-2\im \Sigma_{\text{QC}}(\delta)\right.$ and $f_{\rm res}\omega_{\text P}^2/\omega_{\text{res}}=2\Gamma(0)/(k_0d_z)$. To make the last expression more meaningful, we can approximate $\omega_{\rm res}\approx \omega_0$ and evaluate the right-hand side taking the parameters of the hydrogen atom. This gives the standard formula for the plasma frequency, $\omega_{\text P} = \sqrt{ Nq^2/(m\epsilon_0V)}$, and $f_{\rm res}\approx 0.21$.
Within the limits studied here, we note that the typically phenomenological decay rate $\gamma'$ appearing in the Drude-Lorentz model can be quantitatively connected to specific quantum mechanical processes as encoded in the imaginary part of the self-energy $\Sigma_{\text{QC}}(\delta)$.

\hypersetup{colorlinks=true,
urlcolor  = MidnightBlue
}

\bibliographystyle{my_ieeetr.bst}
\bibliography{main.bib}

\end{document}